\documentclass[aps,twocolumn,PRL,reprint,superscriptaddress]{revtex4}
\usepackage{amsmath}
\usepackage[percent]{overpic}
\usepackage{times}
\usepackage{subfigure}
\usepackage{latexsym}
\usepackage{graphicx}
\usepackage{bm}
\usepackage{amssymb}
\usepackage{anyfontsize}
\usepackage{hyperref}
\usepackage{multirow}
\usepackage{booktabs}
\usepackage{makecell}
\usepackage{natbib}
\begin{document}
\title{Connections between the Open-boundary Spectrum and Generalized Brillouin Zone in Non-Hermitian Systems}
\author{Deguang Wu}
\affiliation{National Laboratory of Solid State Microstructures, Department of Physics, Nanjing University, Nanjing 210093, China}
\author{Jiao Xie}
\affiliation{National Laboratory of Solid State Microstructures, Department of Physics, Nanjing University, Nanjing 210093, China}
\author{Yao Zhou}
\affiliation{National Laboratory of Solid State Microstructures, Department of Physics, Nanjing University, Nanjing 210093, China}
\author{Jin An}
\email{anjin@nju.edu.cn}
\affiliation{National Laboratory of Solid State Microstructures, Department of Physics, Nanjing University, Nanjing 210093, China}
\affiliation{Collaborative Innovation Center of Advanced Microstructures, Nanjing University, Nanjing 210093, China}

\date{\today}

\begin{abstract}
Periodic-boundary spectrum, open-boundary spectrum, as well as the generalized Brillouin zone(GBZ) are three essential properties of a  one-dimensional non-Hermitian system. In this paper we illustrate that the deep connections between them can be revealed by a series of special similar transformations. This viewpoint closely connects the topological geometry of the open-boundary spectrum with the GBZ and provides a new efficient numerical method of calculating them accurately. We further extend these connections to non-Hermitian systems in the symplectic symmetry class. We show that if just the open-boundary features of a non-Hermitian system such as the spectrum and the GBZ, are concerned, the relevant symmetry we should consider is not that of the original system itself, but that of one which has higher symmetry and is related to the original system by a similarity transformation.
\end{abstract}

\maketitle

\textit{Introduction.}
---Recent studies on non-Hermitian systems have revealed dramatic differences from their Hermitian counterparts\cite{hatano1997vortex,hatano1998non,bender1998real,bender2002complex,rudner2009topological,esaki2011edge,hu2011absence,lee2016anomalous,leykam2017edge,shen2018topological,yao2018non,ezawa2019non,okuma2019topological,li2021non,okugawa2021non}. Non-Hermiticity can arise from optical systems with gain and loss\cite{makris2008beam,klaiman2008visualization,guo2009observation,longhi2009bloch,chong2011p,regensburger2012parity,bittner2012p,liertzer2012pump,feng2014single,hodaei2014parity}, open systems with dissipation\cite{carmichael1993quantum,rotter2009non,choi2010quasieigenstate,lin2011unidirectional,lee2014heralded,malzard2015topologically,yamamoto2019theory,li2019observation}, and noninteracting or interacting electron systems with finite-lifetime quasi-particles\cite{zyuzin2018flat,shen2018quantum,yoshida2018non,zhou2018observation,papaj2019nodal,kimura2019chiral}. The central feature of non-Hermitian systems is the sensitivity of the bulk states to boundary conditions, i.e., the non-Hermitian skin effect\cite{yao2018edge,
lee2019anatomy,lee2019hybrid,longhi2019probing,okuma2020topological,yi2020non,kawabata2020higher,liu2020helical,fu2021non,
okuma2021non,sun2021geometric,claes2021skin,guo2021exact}, which results in the breakdown of the conventional bulk-boundary correspondence\cite{kunst2018biorthogonal,herviou2019defining,imura2019generalized,lee2019topological,wang2019non,jin2019bulk,
yang2020non,zhang2020correspondence,zhang2020bulk,wang2020defective,xue2021simple,cao2021non}.
 The efforts of resolving this problem then leads to the establishment of the concept of the GBZ\cite{yao2018edge} and the non-Bloch band theory\cite{yao2018edge,yao2018non,yokomizo2019non,kawabata2020non,yokomizo2021non}. The topological invariants can be redefined on the GBZ instead of BZ to give an explanation of the anomalous bulk-boundary correspondence\cite{yao2018edge,song2019non,deng2019non}. Non-Hermitian topological systems can also exhibit unique features without Hermitian analogs\cite{gong2018topological,yin2018geometrical,jiang2018topological,kawabata2019symmetry,kawabata2019classification,
zhou2019periodic,yokomizo2020topological}, which can be attributed to the complex-valued nature of their spectra. The spectrum of a non-Hermitian system under open-boundary conditions(OBCs) differs greatly from that under periodic-boundary conditions(PBCs)\cite{okuma2020topological,zhang2020correspondence}. The latter can also define a set of winding numbers to characterize the spectrum of the system under the half-infinite boundary conditions\cite{okuma2020topological}. The two spectra, together with the winding numbers, are closely related, and are playing important roles in a non-Hermitian system.

In this paper, we reveal that there are intrinsic connections between the OBC spectrum, the PBC spectrum and the GBZ. We find the bridge which connects them are a series of special similarity transformations. We can thus give an interpretation of the close relation between the geometry of the OBC spectrum such as 3-bifurcation or 4-bifurcation states with the structure of the GBZ. Based on our conclusions, we can also provide a new efficient numerical approach of calculating accurately the OBC spectrum and the GBZ of non-Hermitian systems. Next, we generalize the conclusions on the connections from non-Hermitian systems without symmetry to those with anomalous time-reversal symmetry. We also find that if just the OBC features of a non-Hermitian system such as the OBC spectrum and GBZ, are concerned, the original system together with a series of ones which are related to it by the special similarity transformations, are actually forming an equivalent class which should be treated as a whole, even though these systems may have different symmetries. The open-boundary feature is determined by the symmetry of the system which owns the highest symmetry in the class.

\textit{Correspondence between winding number and the OBC spectrum.}
---We start with a generic quasi-1D tight-binding model whose Hamiltonian reads
\begin{eqnarray}
    \mathcal{H}=\sum_{j}\sum_{m=-M_{2}}^{M_{1}}\sum_{\mu,\nu=1}^{N} T_{m}^{\mu\nu} c_{j+m,\mu}^{\dagger}c_{j,\nu},
    \label{eq1}
\end{eqnarray}
where $\mu$, $\nu$ represent the degrees of freedom including sublattice, spin, orbitals, etc. $T_{m}^{\mu\nu}$ are the hopping amplitudes, and $M_{1}$($M_{2}$) is the range of the hopping to the right(left). Thus, in terms of complex-valued $\beta$, we have the following non-Hermitian Bloch Hamiltonian,
\begin{eqnarray}
    H(\beta)=\sum_{m=-M_{2}}^{M_{1}}{T_{m}{\beta}^{-m}},
    \label{eq2}
\end{eqnarray}
where $T_m$ is the $N\times N$ hopping matrix with $(T_{m})_{\mu\nu}=T_{m}^{\mu\nu}$.

The energy bands $E_{\alpha}(\beta)(\alpha=1,2,\ldots,N)$ of $H(\beta)$ are determined by the characteristic equation det$(E-H(\beta))=0$.
This is an algebraic equation for $\beta$ of $(p+q)$th order, with $p=N{M}_{1}$ and $q=N{M}_{2}$, as the determinant is a polynomial of $E$ and $\beta$, which can be expressed by det$(E-H(\beta))=\prod^{N}_{\alpha=1}(E-{E}_{\alpha}(\beta))=a_{-p}\frac{1}{{\beta}^{p}}+\ldots+a_{q}{\beta}^{q}$. For a fixed eigenenergy $E$, the $p+q$ solutions ${\beta}_{i}(E)$ are assumed to be numbered as $|{\beta}_{1}(E)|\leq|{\beta}_{2}(E)|\leq \ldots \leq |{\beta}_{p+q}(E)|$. For a non-Hermitian system without any symmetry, a state belonging to the OBC continuum bands must have to obey the condition $|{\beta}_{p}(E)|=|{\beta}_{p+1}(E)|$\cite{yao2018non,yokomizo2019non}, and the trajectories of ${\beta}_{p}$ and ${\beta}_{p+1}$ give the GBZ.

To illustrate the connections between the OBC spectrum and the GBZ in non-Hermitian systems, we study a series of systems described by Hamiltonians $\mathcal{H}_{\rho}$, which are related to the matrix Hamiltonian of $\mathcal{H}$ by a special similarity transformation,
\begin{eqnarray}
    \mathcal{H}_{\rho}\equiv S^{-1}\mathcal{H}S,
    \label{eq3}
\end{eqnarray}
where $S={\rm{diag}}\left\{1,\rho,\rho^{2},\ldots,\rho^{L}\right\}\bigotimes \mathbb{I}$ is a block-diagonal matrix, with $\mathbb{I}$ the $N\times N$ identity matrix and $L$ the system size.

According to Eq.(2), it is clear that ${H}_{\rho}(\beta)=H(\rho\beta)$, as the hopping matrix $T_{m}$ becomes $T_{m}\rho^{-m}$ under the similarity transformation. For open systems, since ${\rm{det}}(S^{-1}\mathcal{H}S-E)={\rm{det}}(\mathcal{H}-E)=0$, $\mathcal{H}_{\rho}$ and $\mathcal{H}$ share identical OBC spectrum. The unit circle $|\beta|=1$ on the complex $\beta$-plane is the BZ of $H(\beta)$. The map of this circle onto the complex $E$-plane by $E_{\alpha}(\beta)$ are $N$ curves $E_{\alpha}(|\beta|=1)(\alpha=1,2,\ldots,N)$, which form the PBC spectrum of $H(\beta)$ and is denoted for simplicity as $E(|\beta|=1)$.  Because the BZ of the system described by ${H}_{\rho}(\beta)$ exactly corresponds to $|\beta|=\rho$ of the original system described by $H(\beta)$, the PBC spectrum of ${H}_{\rho}(\beta)$ can be faithfully given by $E(|\beta|=\rho)$. For a definite $E$, if we denote the solutions of det$(H(\beta)-E)=0$ as ${\beta}_{i}(E)$, then the solutions of det$(H(\rho\beta)-E)=0$ must be $\beta_{i}(E)/\rho$, where $i=$1, 2,$\cdot\cdot\cdot, p+q$. Thus for any positive-valued $\rho$, ${H}_{\rho}(\beta)$ shares with $H(\beta)$ the same condition $|{\beta}_{p}(E)|=|{\beta}_{p+1}(E)|$ for the OBC continuum bands, resulting in that ${H}_{\rho}(\beta)$ has exactly the same shape of GBZ to that of ${H}(\beta)$, but with a scale factor $\rho^{-1}$.

Due to the multi-valued nature of $E_{\alpha}(\beta)$, each curve of $E(|\beta|=\rho)$ is not necessarily closed but all the curves must always form some closed loops with their number being $\mathcal{N}_{\rho}\leq N$. The $\mathcal{N}_{\rho}$ closed loops divide the whole $E$-plane into several areas. Each area can be characterized by a winding number which is the sum of the phase winding of each loop around any point within this area. For any such point $E$, this winding number can be formally defined by
\begin{align}
    w_{\rho}(E)
    =&\frac{1}{2\pi\mathrm{i}}{\oint}_{|\beta|=1}d\beta \frac{d}{d\beta}{\rm{det}}(H_{\rho}(\beta)-E)\nonumber\\
    =&\frac{1}{2\pi\mathrm{i}}{\oint}_{|\beta|=\rho}d\beta \frac{d}{d\beta}{\rm{ln}}({\rm{det}}(H(\beta)-E))\nonumber\\
    =&\sum_{i=1}^{p+q}\frac{1}{2\pi\mathrm{i}}\oint_{|\beta|=\rho}d\beta{\rm{ln}}(\beta-\beta_{i}(E))-p\nonumber\\
    =&N_{\rho}(E)-p,
    \label{eq4}
\end{align}
where $N_{\rho}(E)$ is the number of $\beta_{i}(E)$ inside the circle $|\beta|=\rho$.

We consider the crossing points of these closed loops, which can be self-intersection ones between the same loop or intersection ones between different loops. One of our central results is a theorem: For a non-Hermitian system without any symmetry, for a definite $\rho$, if there exists a crossing point $E_{c}$ which is on the boundary between two areas with winding number $w_{\rho}=+1$ and $w_{\rho}=-1$ respectively, then $E_{c}$ belongs to the OBC spectrum. Conversely, for any ``normal'' point $E_{s}$ on the OBC spectrum, there must be one and only one $\rho$ so that the PBC spectrum $E(|\beta|=\rho)$ has a crossing point exactly at $E_{s}$, with their two adjacent areas possessing winding numbers $\pm1$.
\begin{figure}
 \centering
    \includegraphics[width=0.48\textwidth]{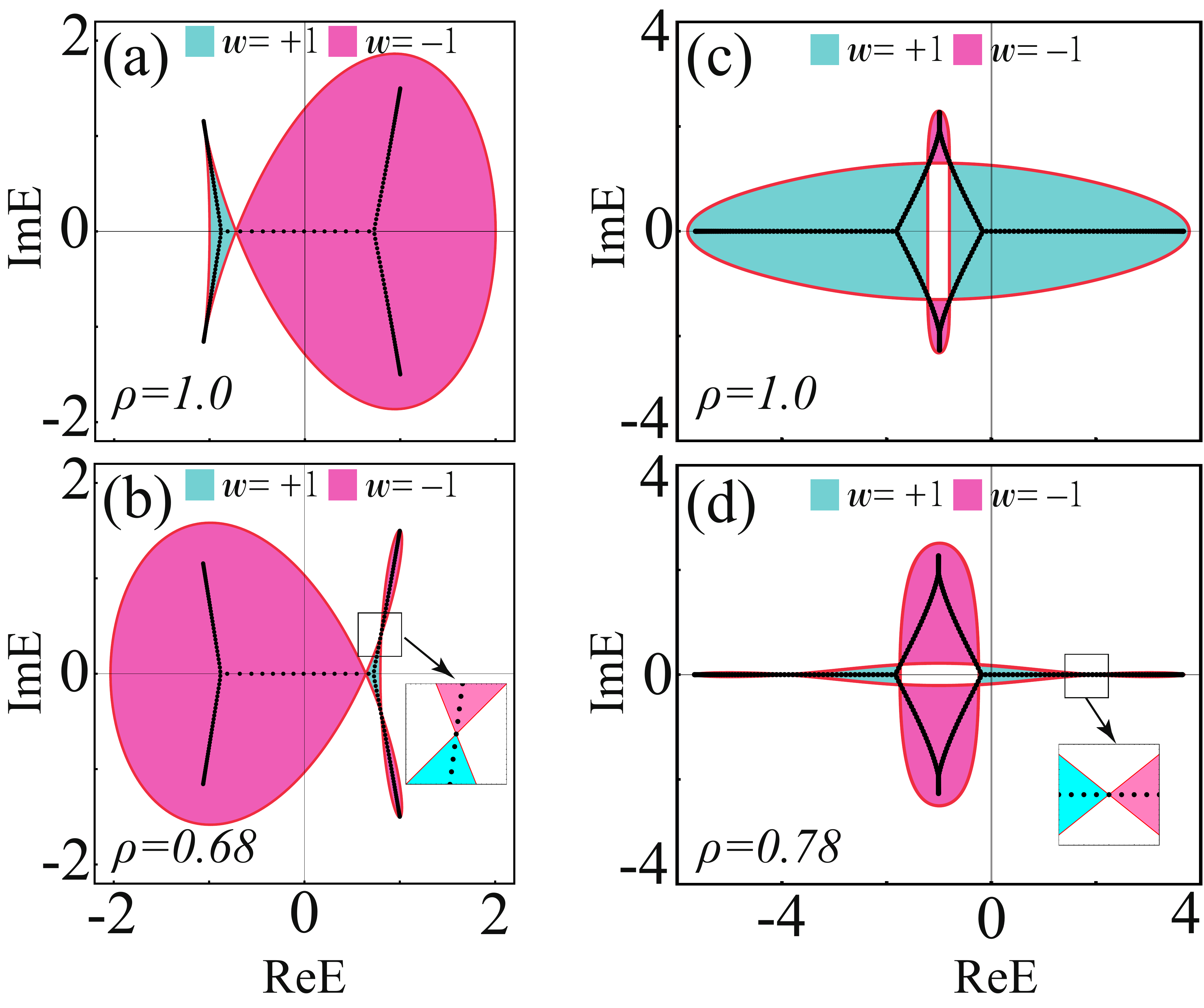}
    \caption{Connections between the PBC spectrum and OBC spectrum via winding numbers.
    The solid(dotted) lines denote the spectra under PBCs(OBCs).
    (a)-(b) are for a single-band model with different $\rho$, where the nonzero parameters are $T_{-2}=-1$, $T_{-1}=\mathrm{i}$, $T_1=-0.5\mathrm{i}$, $T_2=0.5$, while (c)-(d) are for a two-band model, where the nonzero parameters are $T_{-1}=\mathbb{I}+2 \sigma_{3}$, $T_{0}=-\mathbb{I}+\mathrm{i} \sigma_{2}$, $T_{1}=2\mathbb{I}+\sigma_{3}$, with $\sigma_{i}(i=1,2,3)$ Pauli matrices.
    The colored areas are possessing winding number $w_{\rho}(E)=\pm1$.}
    \label{fig1}
\end{figure}

In Fig.1 we illustrate the essential points of the above result by using both a single-band and two-band model as examples. In the single-band model, even one closed loop give multiple self-crossing points which are located on the OBC spectrum, while in the two-band model, for a definite $\rho$, the PBC spectrum $E(|\beta|=\rho)$ give simultaneously the self-crossing points and intersection points between the two loops, which also belong to the corresponding OBC spectrum.

\begin{figure}
    \centering
    \includegraphics[width=0.48\textwidth]{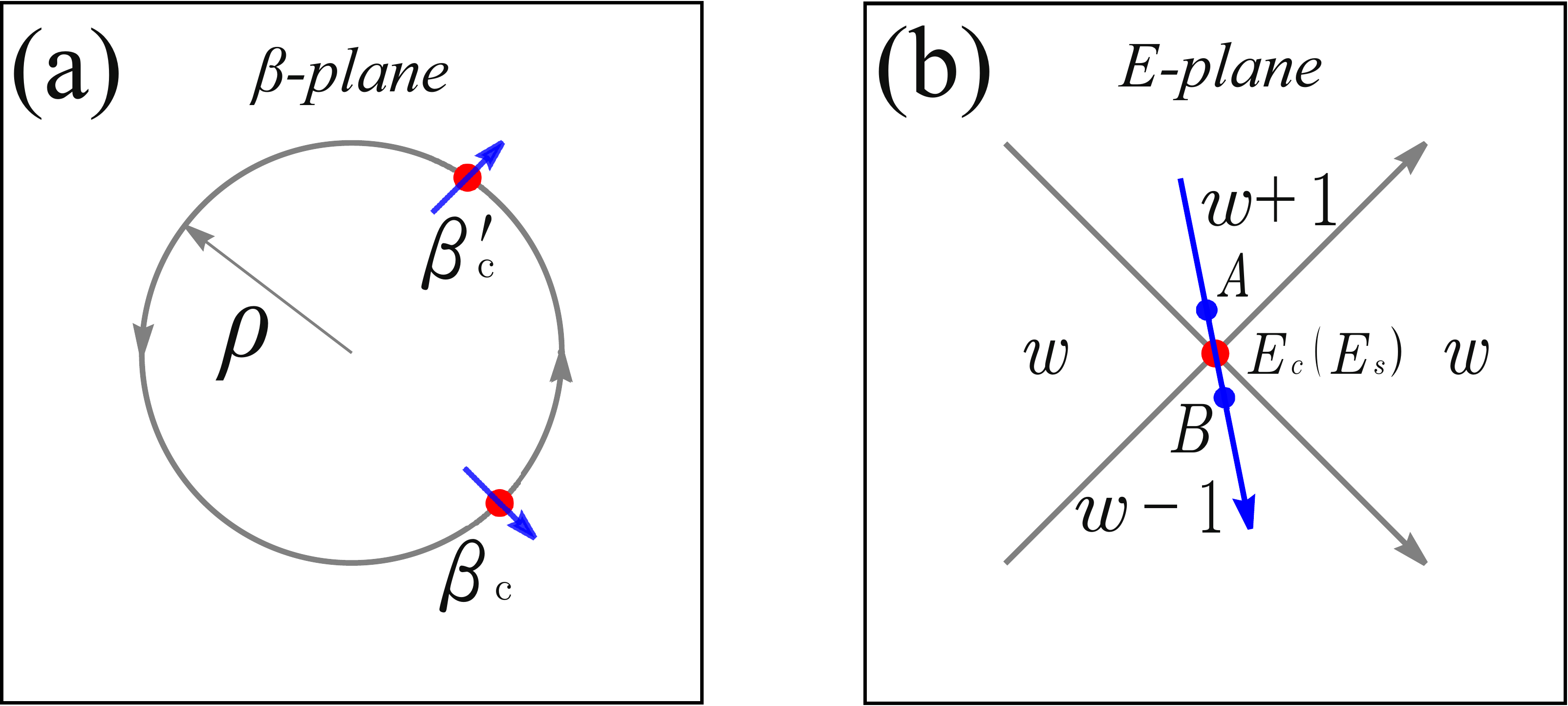}
    \caption{Illustration of the theorem: (a) $|\beta|=\rho$ and (b) a crossing point $E_{c}$ of the corresponding PBC spectrum $E(|\beta|=\rho)$, where the four areas adjacent to $E_{c}$ have been marked by their winding numbers. The spectrum on the $E$-plane is the image of the circle on the $\beta$-plane, under the mapping $E(\beta)$. The two solid dots $\beta_{c}$, $\beta^{'}_{c}$ on the circle in (a) are the preimages of $E_{c}$ in (b), and accordingly, the two small arcs passing through the two points corresponds to the trajectory of from A to B in (b).  }
\label{fig2}
\end{figure}

Now we give the proof. Let $E_{c}$ be a crossing point of a PBC spectrum $E(|\beta|=\rho)$. This crossing point means that there must be two different $\beta_{c}$ and $\beta^{'}_{c}$ on the circle $|\beta|=\rho$( see Fig.2(a)), satisfying $E_{\alpha}(\beta_{c})=E_{\gamma}(\beta^{'}_{c})=E_{c}$ for two different bands $\alpha$ and $\gamma$, or $E_{\alpha}(\beta_{c})=E_{\alpha}(\beta^{'}_{c})=E_{c}$ for the same band $\alpha$. Adjacent to $E_{c}$, there are four areas with winding numbers $(w+1)/w/(w-1)/w$ respectively, as shown in Fig~\ref{fig2}.(b). We have to prove that if $w=0$, $E_{c}$ is on the OBC spectrum. Consider a small curve starting from point $A$ to point B, which is passing through $E_{c}$. The area where $A/B$ sits possesses winding number $(w+1)/(w-1)$. Due to Eq.(~\ref{eq4}), the number of $\beta_{i}(E_{A/B})$ inside the circle $|\beta|=\rho$ should be $(p+w+1)/(p+w-1)$. If $A$ and $B$ are sufficiently close to $E_{c}$, among the $p+q$ image curves of the map $\beta_{i}(E)$ of the trajectory $E_{A}\rightarrow E_{c}\rightarrow E_{B}$ onto the $\beta$-plane, only two small curves would intersect the circle $|\beta|=\rho$ at $\beta=\beta_{c}$ and $\beta=\beta^{'}_{c}$. Moreover, these two curves must be pointing outwards, as shown in Fig~\ref{fig2}.(a), in order to reduce the winding number by two. Therefore, the number of $\beta_{i}(E_{c})$ inside the circle should be $p+w-1$. If $w=0$, then $\beta_{c}$ and $\beta^{'}_{c}$ must be the $p$th and $(p+1)$th $\beta$ solutions of $E_{c}$ and then their equal absolute value means that $E_{c}$ is on the OBC spectrum.

Conversely, consider a point $E_{s}$ on the OBC spectrum. This $E_{s}$ generally corresponds to two different $\beta_{p}(E_{s})$, $\beta_{p+1}(E_{s})$ on the GBZ. Since $|\beta_{p}(E_{s})|=|\beta_{p+1}(E_{s})|=\rho_{s}$, the PBC spectrum $E(|\beta|=\rho_{s})$ would inevitable form a crossing point at $E_{s}$. Obviously, the number of $\beta_{i}(E_{s})$ inside the circle $|\beta|=\rho_{s}$ is $p-1$, and for the four areas adjacent to $E_{s}$, similar to the situation as shown in Fig.2(b), their winding numbers must be $(+1)/0/(-1)/0$. Thus we complete the proof.

Therefore, any crossing point of a PBC spectrum of $H_{\rho}(\beta)$ with $\pm1$ winding numbers in its two adjacent areas must be on the OBC spectrum. This theorem thus provides an accurate and effective new method to numerically obtain the OBC spectrum as well as the GBZ for 1D non-Hermitian systems\cite{Supp1}. Compared with the previous works\cite{yokomizo2019non,yang2020non}, this numerical method is much easier to handle.

Obviously, not every crossing point is on the OBC spectrum, because some crossing points are not adjacent to the areas with winding numbers $\pm1$. These energy states $E$ generally require $|\beta_{i}(E)|=|\beta_{i+1}(E)|$ with $i\neq p$, and thus the two $\beta$ values should be on the auxiliary GBZ introduced in Ref.\cite{yang2020non} in obtaining the GBZ.

An exception to the theorem is when the GBZ is a perfect circle $|\beta|=r$. This can happen even in multi-band models. In this particular situation, the PBC spectrum $E(|\beta|=r)$ covers the whole OBC spectrum, and the other PBC spectra $E(|\beta|=\rho)$ with $\rho\neq r$ give no contributions to the OBC spectrum\cite{Supp2}.

\textit{N-bifurcation states, topological geometry of the OBC spectrum and their connection with the GBZ.}
---According to the theorem, nearly any state on the OBC spectrum can be a crossing point of a PBC spectrum for a certain $\rho$.
These are the ``normal'' points mentioned in the theorem, which nearly constitute the whole OBC spectrum and are characterized by their unique feature that each has two and only two extending directions along the spectrum. However, there always exist several other energy states on the OBC spectrum, which are the 3-bifurcation or 4-bifurcation states\cite{Supp3}(see the solid dots in Fig.3(a) and (c)), and end-point states. These particular states can also be the crossing points of some PBC spectra, but with different winding number distributions in their adjacent areas. Take a 3-bifurcation state $E$ as an example. We find it obeys,
\begin{align}
    |\beta_{p-1}(E)|=|\beta_{p}(E)|=|\beta_{p+1}(E)|,\nonumber\\
    {\rm{or}}\qquad |\beta_{p}(E)|=|\beta_{p+1}(E)|=|\beta_{p+2}(E)|,
    \label{eq5}
\end{align}
i.e., each 3-bifurcation state has not two, but three $\beta$ solutions on the GBZ. More generally, a $n$-bifurcation state $E$ must have exactly $n$ $\beta$ solutions on the GBZ. In this opinion, all the energy states can be viewed in an universal way. The end points on the OBC spectrum can be viewed as ``1-bifurcation'' ones and the ``normal'' points as the 2-bifurcation ones. The single $\beta$ on the GBZ corresponding to an end-point state $E$ is actually a doubly degenerate $\beta$ satisfying $\beta_{p}=\beta_{p+1}$, as discussed in Ref.\cite{yang2020non}. These behavior are exhibited in Fig.3.
In the one-band model with $p=q=4$, for the 3-bifurcation states, we have $|\beta_{3}|=|\beta_{4}|=|\beta_{5}|$, and for the 4-bifurcation state, we have $|\beta_{4}|=|\beta_{5}|=|\beta_{6}|=|\beta_{7}|$. In the two-band model with $p=q=2$, for the four 3-bifurcation states, we have $|\beta_{1}|=|\beta_{2}|=|\beta_{3}|$ for $A_{1}$ and $A_{2}$, while $|\beta_{2}|=|\beta_{3}|=|\beta_{4}|$ for $B_{1}$ and $B_{2}$.

To prove Eq.(5), for a 3-bifurcation state $E$, assume its counterpart on the GBZ is a set of $n$ $\beta_{i}(E)$ points, where $i=j+1,j+2,...j+n$, obeying $|\beta_{j+1}(E)|=|\beta_{j+2}(E)|=...=|\beta_{j+n}(E)|$, with $j<p$ and $j+n>p$. Consider three small arcs along the three branches connecting to $E$ (see Fig.3(c)). Because any other state on the arcs is a normal one which corresponds to two different $\beta$s on the GBZ, the three small arcs must be mapped onto the GBZ to the six small arcs departing from the $n$ $\beta_{i}(E)$. But departing from any $\beta_{i}(E)$, there exist only two extending directions to accommodate two small arcs. Then $n$ must be three and we complete the proof.
\begin{figure}
   \centering
    \includegraphics[width=0.48\textwidth]{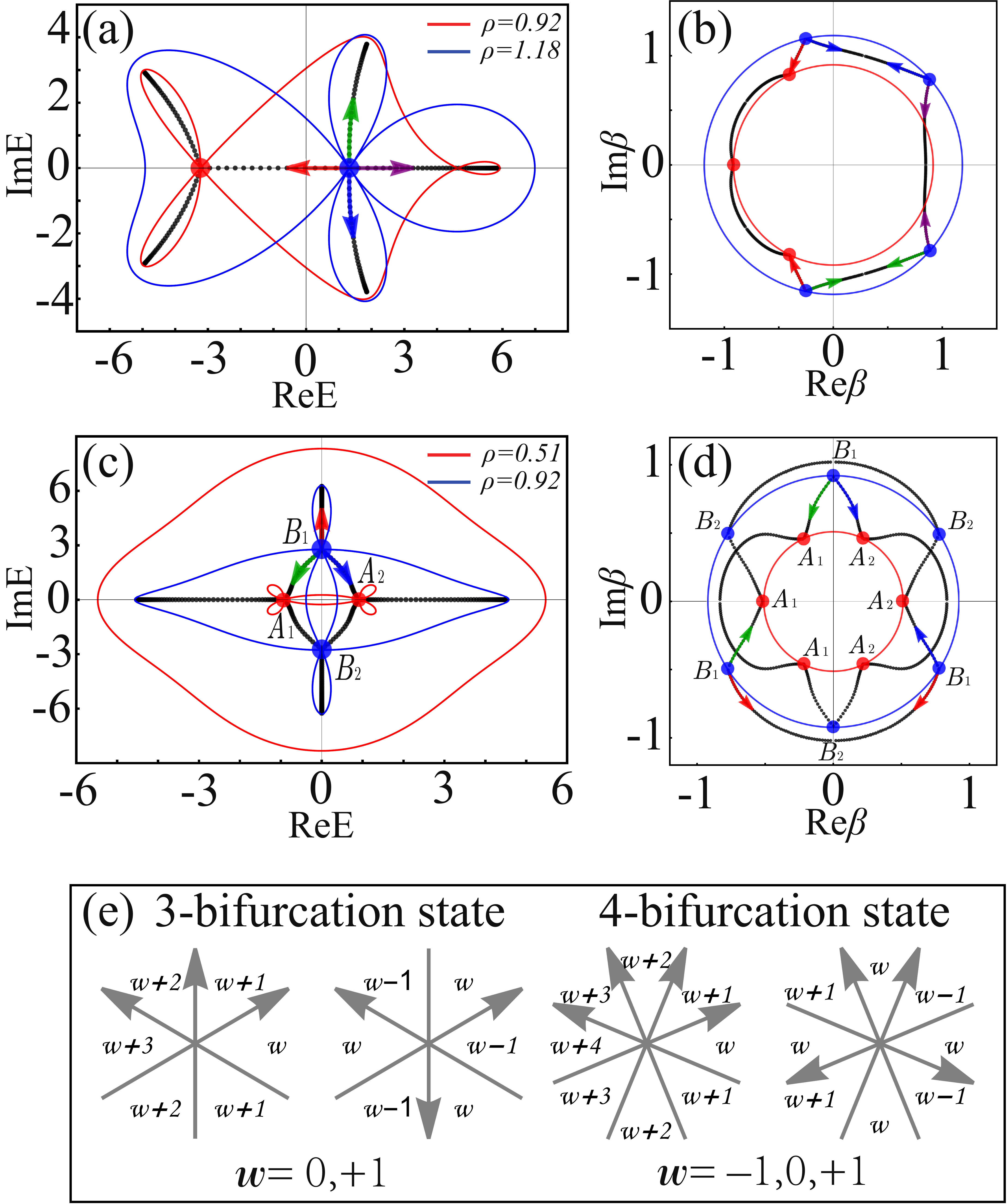}
    \caption{N-bifurcation states on the OBC spectrum and the GBZ. (a) and (c): 3-bifurcation and 4-bifurcation states, on the OBC spectrum, together with the PBC spectra passing through them.
    (b) and (d): GBZ, where the solid circles are the corresponding three(four) $\beta$s for the 3(4)-bifurcation states. Here (a)-(b) are for a single-band model with the nonzero parameters $T_{-4}=0.8$, $T_{\pm{3}}=\mp{1}$, $T_{\pm{2}}=\pm{1}$, $T_{-1}=2$, $T_{1}=3$, $T_{4}=0.3$, while (c)-(d) are for a two-band model with the nonzero parameters are $T_{-1}=1.5\mathbb{I}+\sigma_{1}+0.5\sigma_{3}$, $T_{0}=\mathrm{i}\sigma_{2}$, $T_{1}=0.5\mathbb{I}+\sigma_{1}-2.5\sigma_{3}$,where the PBC spectrum passing through the 3-bifurcation state $A_{1}(B_{1})$ is also passing through $A_{2}(B_{2})$. The dotted loops in (b) and (d) are the circles $|\beta|=\rho$ corresponding to the PBC spectra in (a) and (c). (e) All possible winding number distributions adjacent to a 3-bifurcation or 4-bifurcation state.}
\label{fig3}
\end{figure}
\par For a generic $n$-bifurcation state $E$, since there are $n$ successive $\beta_{i}(E)$ solutions with the same absolute value located on the GBZ, the PBC spectrum $E(|\beta|=\rho_{0})$ is passing through $E$ exactly $n$ times, where $\rho_{0}=|\beta_{p}(E)|$. The local areas adjacent to $E$ would be divided into $2n$ parts. A $n$-bifurcation state generally has relatively definite winding number distributions in these adjacent areas. For a 3-bifurcation or 4-bifurcation energy state, all possible winding number distributions are shown explicitly in Fig.3(e). The $n$ $\beta_{i}(E)$ solutions on the GBZ can also be viewed in another way: They are the intersection points between the circle $|\beta|=\rho$ and the GBZ, as seen in Fig.3(b) and Fig.3(d). So starting with a circle $|\beta|=\rho$, with $\rho$ increasing gradually from $0$ to infinity, one has a set of intersection points for each $\rho$ with the GBZ. Each set determines one or several energy states $E$ on the OBC spectrum and the number of the points in the set indicates the topological geometry of this state: One only means an end-point state, two means a ``normal'' state or two end-point states, and three means a 3-bifurcation state, or a ``normal'' state together with an end-point state, or three end-point states, etc. Thus the $\rho$ parameter of the PBC spectrum having crossing points on the open spectrum has actually a finite range: $|\beta|_{\rm{min}}\leq\rho\leq|\beta|_{\rm{max}}$, where $|\beta|_{\rm{max}/\rm{min}}$ is the maximum/minimum value of $|\beta|$ on the GBZ. Now we can also give the non-Hermitian skin effect a new viewpoint. Under OBCs, any state belonging to the continuum bands generally will be localized at end. However, if we assign each state $E$ on the OBC spectrum an unique open system described by $H_{\rho}(\beta)$, with $\rho=|\beta_{p}(E)|$, then this state with respect to $H_{\rho}(\beta)$ must be extended, since its $\beta$ solutions $\beta^{'}_{i}(E)=\beta_{i}(E)/\rho$ obey: $|\beta^{'}_{p}(E)|=|\beta^{'}_{p+1}(E)|=1$. Thus from this opinion, the whole OBC spectrum of the continuum bands can still be viewed as being extended without skin effect, not for the original system $H(\beta)$, but for a series of representatives $H_{\rho}(\beta)$ with $\rho=|\beta_{p}(E)|$.

\textit{Extension to systems with anomalous time-reversal symmetry.}
---Now we extend our results to systems with symmetry. We focus on the non-Hermitian systems in the symplectic symmetry class, which has the anomalous time-reversal symmetry\cite{yi2020non,kawabata2020non,kawabata2019symmetry} defined by: $U_T \mathcal{H}^t U^{-1}_{T}=\mathcal{H}$, where $U_T$ is a unitary matrix satisfying $U_T {U_T}^*=-1$. This symmetry indicates $U_{T}H^{t}(\beta^{-1})U^{-1}_{T}=H(\beta)$, so the standard non-Bloch band theory breaks down and should be modified according to Ref.\cite{kawabata2020non}. The characteristic equation det$[H(\beta)-E]=0$ has $2p$ numbered solutions $ \beta_{i}(E)$ in total( $ p=q=NM_{1}$), obeying $\beta_{p+i}(E)=\beta_{p+1-i}^{-1}(E) $ due to Kramers degeneracy\cite{kawabata2020non}, which results in a new condition for the GBZ: $|\beta_{p-1}(E)|=|\beta_{p}(E)|$ or $|\beta_{p+1}(E)|=|\beta_{p+2}(E)|$\cite{Supp4}.

Analogous discussion leads us to a similar conclusion for a non-Hermitian system in the symplectic symmetry class: The crossing points of the PBC spectrum $E(|\beta|=\rho)$ with its two adjacent areas possessing winding numbers $\pm2/0$ belong to the OBC spectrum. Conversely, for any ``normal'' point $E_{s}$ on the OBC spectrum, these must exist two $\rho_{1}$ and $\rho_{2}$ with $\rho_{1}\rho_{2}=1$ and $\rho_{1}\leq\rho_{2}$ so that the PBC spectrum $E(|\beta|=\rho_{1})$ ( $E(|\beta|=\rho_{2})$) has a crossing point exactly at $E_{s}$, with its two adjacent areas possessing winding numbers $-2/0$(+2/0).

\begin{figure}
    \centering
    \includegraphics[width=0.48\textwidth]{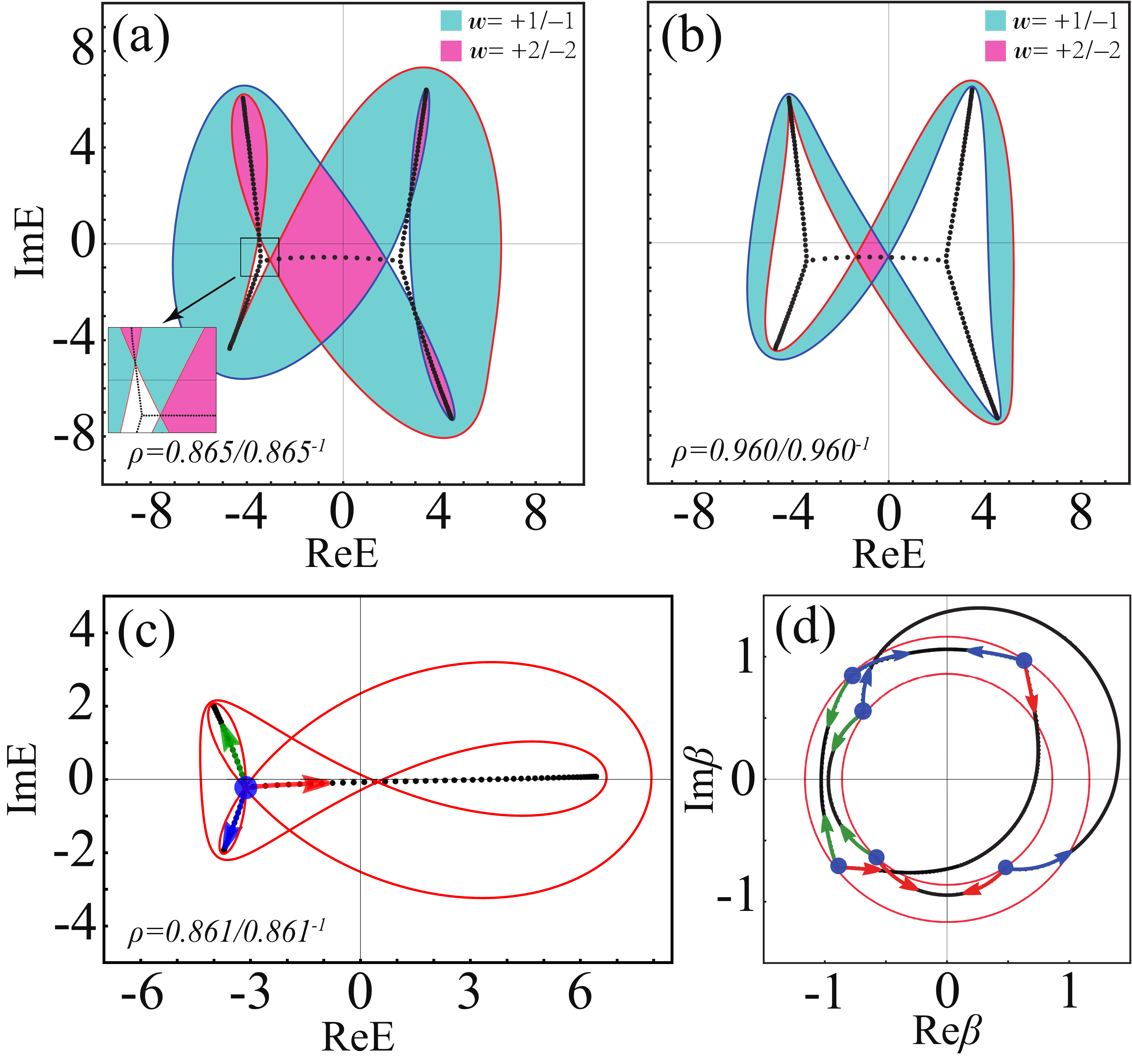}
    \caption{A non-Hermitian system in the symplectic symmetry class. (a) OBC spectrum, PBC spectra for a two-band model with the non-zero parameters $T_{\pm{1}}=2\mathbb{I}\pm(1+1.5 \mathrm{i})\sigma_{1}\pm\sigma_{2}\pm2 \mathrm{i}\sigma_{3}$, $T_{\pm{2}}=3 \mathrm{i}\mathbb{I}\pm(-0.6+0.2 \mathrm{i})\sigma_{1}\pm\sigma_{2}\pm0.8\sigma_{3}$. Here the crossing points of the PBC spectrum in (a) have two adjacent areas with winding number $0$ or $\pm2$ and so they belong to the OBC spectrum. (c) A 3-bifurcation state on the OBC spectrum and the PBC spectra passing through it, and (d) its counterpart on the GBZ, where the nonzero parameters are $T_{\pm{1}}=2\mathbb{I}\pm1.5 \mathrm{i}\sigma_{1}\pm1.5 \mathrm{i}\sigma_{2}\pm\sigma_{3}$, $T_{\pm{2}}=\mathrm{i}\mathbb{I}\pm0.7 \mathrm{i}\sigma_{1}\pm0.7 \mathrm{i}\sigma_{2}\pm0.5\sigma_{3}$. In both models, $U_{T}=\sigma_{2}$.}
\label{fig4}
\end{figure}
Now we give our explanation. Due to the symmetry $U_{T}H^{t}(\beta^{-1})U^{-1}_{T}=H(\beta)$, the $\mathcal{N}_{\rho}$ closed loops of the PBC spectrum $E(|\beta|=\rho)$ exactly coincide with those of the PBC spectrum $E(|\beta|=\rho^{-1})$, but the two sets of loops are pointing in opposite directions. So each state on the OBC spectrum will be experienced twice by the PBC ones. This also indicates that any area on the $E-$plane has opposite winding number with respect to $E(|\beta|=\rho)$ and $E(|\beta|=\rho^{-1})$. Since here a ``normal" state on the OBC spectrum corresponds to four $\beta$s on the GBZ, an end-point state should have two degenerate points on the GBZ obeying: $\beta_{p-1}=\beta_{p}$ or $\beta_{p}=\beta_{p+1}$. The counterpart of a 3-bifurcation state should have two sets of $\beta$ points on the GBZ, with each set being composed of the three of them, which obey:
\begin{align}
    |\beta_{p-2}(E)|=|\beta_{p-1}(E)|=|\beta_{p}(E)|,\nonumber\\
    {\rm{or}}\qquad |\beta_{p+1}(E)|=|\beta_{p+2}(E)|=|\beta_{p+3}(E)|.
    \label{eq6}
\end{align}
In this case, there still exist two PBC spectra for $\rho_{1}$ and $\rho_{2}$ with $\rho_{1}\rho_{2}=1$, each of which is passing through the 3-bifurcation state three times and is overlapping exactly with each other. We illustrate these results by a two-band model in Fig.4, where $p=4$, and the 3-bifurcation state obeys: $|\beta_{2}|=|\beta_{3}|=|\beta_{4}|$, or $|\beta_{5}|=|\beta_{6}|=|\beta_{7}|$.

It should be noted that $H_{\rho}(\beta)$ with $\rho\neq 1$ generally have no symmetry, but they share identical OBC spectrum with $H(\beta)$ since they are related to each other by similarity transformations. What's remarkable is that $H_{\rho}(\beta)$ share with $H(\beta)$ the same GBZ conditions, even if they possess different symmetries. This is because, as mentioned before, for the same state with eigenenergy $E$, the $\beta_{i}(E)$ solutions of det$(E-H_{\rho}(\beta))=0$ is proportional to those of det$(E-H(\beta))=0$. To make this point more clear, we consider a class of Hamiltonians related by the special similarity transformations, $\mathcal{C}_{H}=\{H_{\rho}(\beta)=H(\rho\beta)\mid 0<\rho<\infty\}$. If the original system we start with is described by $H(\beta)$ which does not have any symmetry, and simultaneously, one member $H_{\rho}(\beta)$ of the class has certain symmetry which leads to a modified GBZ conditions, then $H(\beta)$ must have the same GBZ conditions. Therefore, if only the OBC features of a non-Hermitian system such as the spectrum and the GBZ, are concerned, the relevant symmetry we should consider is not that of the original system $H(\beta)$ itself, but the one belonging to $\mathcal{C}_{H}$ which has the highest symmetry. So to extract the properties of an open non-Hermitian systems, the class $\mathcal{C}_{H}$ should be treated as a whole. This viewpoint could be extended straightforwardly to higher dimensions.

This work is supported by NSFC under Grants No.11874202.

\nocite{*}

\bibliographystyle{unsrt}
\bibliography{maintext}

\begin{thebibliography}{10}

\bibitem{hatano1997vortex}
Naomichi Hatano and David~R Nelson.
\newblock Vortex pinning and non-hermitian quantum mechanics.
\newblock {\em Physical Review B}, 56(14):8651, 1997.

\bibitem{hatano1998non}
Naomichi Hatano and David~R Nelson.
\newblock Non-hermitian delocalization and eigenfunctions.
\newblock {\em Physical Review B}, 58(13):8384, 1998.

\bibitem{bender1998real}
Carl~M Bender and Stefan Boettcher.
\newblock Real spectra in non-hermitian hamiltonians having p t symmetry.
\newblock {\em Physical Review Letters}, 80(24):5243, 1998.

\bibitem{bender2002complex}
Carl~M Bender, Dorje~C Brody, and Hugh~F Jones.
\newblock Complex extension of quantum mechanics.
\newblock {\em Physical Review Letters}, 89(27):270401, 2002.

\bibitem{rudner2009topological}
Mark~S Rudner and LS~Levitov.
\newblock Topological transition in a non-hermitian quantum walk.
\newblock {\em Physical Review Letters}, 102(6):065703, 2009.

\bibitem{esaki2011edge}
Kenta Esaki, Masatoshi Sato, Kazuki Hasebe, and Mahito Kohmoto.
\newblock Edge states and topological phases in non-hermitian systems.
\newblock {\em Physical Review B}, 84(20):205128, 2011.

\bibitem{hu2011absence}
Yi~Chen Hu and Taylor~L Hughes.
\newblock Absence of topological insulator phases in non-hermitian p
  t-symmetric hamiltonians.
\newblock {\em Physical Review B}, 84(15):153101, 2011.

\bibitem{lee2016anomalous}
Tony~E Lee.
\newblock Anomalous edge state in a non-hermitian lattice.
\newblock {\em Physical Review Letters}, 116(13):133903, 2016.

\bibitem{leykam2017edge}
Daniel Leykam, Konstantin~Y Bliokh, Chunli Huang, Yi~Dong Chong, and Franco
  Nori.
\newblock Edge modes, degeneracies, and topological numbers in non-hermitian
  systems.
\newblock {\em Physical Review Letters}, 118(4):040401, 2017.

\bibitem{shen2018topological}
Huitao Shen, Bo~Zhen, and Liang Fu.
\newblock Topological band theory for non-hermitian hamiltonians.
\newblock {\em Physical Review Letters}, 120(14):146402, 2018.

\bibitem{yao2018non}
Shunyu Yao, Fei Song, and Zhong Wang.
\newblock Non-hermitian chern bands.
\newblock {\em Physical Review Letters}, 121(13):136802, 2018.

\bibitem{ezawa2019non}
Motohiko Ezawa.
\newblock Non-hermitian boundary and interface states in nonreciprocal
  higher-order topological metals and electrical circuits.
\newblock {\em Physical Review B}, 99(12):121411, 2019.

\bibitem{okuma2019topological}
Nobuyuki Okuma and Masatoshi Sato.
\newblock Topological phase transition driven by infinitesimal instability:
  Majorana fermions in non-hermitian spintronics.
\newblock {\em Physical Review Letters}, 123(9):097701, 2019.

\bibitem{li2021non}
Tianyu Li, Jia-Zheng Sun, Yong-Sheng Zhang, and Wei Yi.
\newblock Non-bloch quench dynamics.
\newblock {\em Physical Review Research}, 3(2):023022, 2021.

\bibitem{okugawa2021non}
Ryo Okugawa, Ryo Takahashi, and Kazuki Yokomizo.
\newblock Non-hermitian band topology with generalized inversion symmetry.
\newblock {\em Physical Review B}, 103(20):205205, 2021.

\bibitem{makris2008beam}
Konstantinos~G Makris, R~El-Ganainy, DN~Christodoulides, and Ziad~H Musslimani.
\newblock Beam dynamics in p t symmetric optical lattices.
\newblock {\em Physical Review Letters}, 100(10):103904, 2008.

\bibitem{klaiman2008visualization}
Shachar Klaiman, Uwe G{\"u}nther, and Nimrod Moiseyev.
\newblock Visualization of branch points in p t-symmetric waveguides.
\newblock {\em Physical Review Letters}, 101(8):080402, 2008.

\bibitem{guo2009observation}
A~Guo, GJ~Salamo, D~Duchesne, R~Morandotti, M~Volatier-Ravat, V~Aimez,
  GA~Siviloglou, and DN~Christodoulides.
\newblock Observation of p t-symmetry breaking in complex optical potentials.
\newblock {\em Physical Review Letters}, 103(9):093902, 2009.

\bibitem{longhi2009bloch}
Stefano Longhi.
\newblock Bloch oscillations in complex crystals with p t symmetry.
\newblock {\em Physical Review Letters}, 103(12):123601, 2009.

\bibitem{chong2011p}
YD~Chong, Li~Ge, and A~Douglas Stone.
\newblock P t-symmetry breaking and laser-absorber modes in optical scattering
  systems.
\newblock {\em Physical Review Letters}, 106(9):093902, 2011.

\bibitem{regensburger2012parity}
Alois Regensburger, Christoph Bersch, Mohammad-Ali Miri, Georgy Onishchukov,
  Demetrios~N Christodoulides, and Ulf Peschel.
\newblock Parity--time synthetic photonic lattices.
\newblock {\em Nature}, 488(7410):167--171, 2012.

\bibitem{bittner2012p}
S~Bittner, B~Dietz, Uwe G{\"u}nther, HL~Harney, M~Miski-Oglu, A~Richter, and
  F~Sch{\"a}fer.
\newblock P t symmetry and spontaneous symmetry breaking in a microwave
  billiard.
\newblock {\em Physical Review Letters}, 108(2):024101, 2012.

\bibitem{liertzer2012pump}
M~Liertzer, Li~Ge, A~Cerjan, AD~Stone, Hakan~E T{\"u}reci, and S~Rotter.
\newblock Pump-induced exceptional points in lasers.
\newblock {\em Physical Review Letters}, 108(17):173901, 2012.

\bibitem{feng2014single}
Liang Feng, Zi~Jing Wong, Ren-Min Ma, Yuan Wang, and Xiang Zhang.
\newblock Single-mode laser by parity-time symmetry breaking.
\newblock {\em Science}, 346(6212):972--975, 2014.

\bibitem{hodaei2014parity}
Hossein Hodaei, Mohammad-Ali Miri, Matthias Heinrich, Demetrios~N
  Christodoulides, and Mercedeh Khajavikhan.
\newblock Parity-time--symmetric microring lasers.
\newblock {\em Science}, 346(6212):975--978, 2014.

\bibitem{carmichael1993quantum}
HJ~Carmichael.
\newblock Quantum trajectory theory for cascaded open systems.
\newblock {\em Physical Review Letters}, 70(15):2273, 1993.

\bibitem{rotter2009non}
Ingrid Rotter.
\newblock A non-hermitian hamilton operator and the physics of open quantum
  systems.
\newblock {\em Journal of Physics A: Mathematical and Theoretical},
  42(15):153001, 2009.

\bibitem{choi2010quasieigenstate}
Youngwoon Choi, Sungsam Kang, Sooin Lim, Wookrae Kim, Jung-Ryul Kim, Jai-Hyung
  Lee, and Kyungwon An.
\newblock Quasieigenstate coalescence in an atom-cavity quantum composite.
\newblock {\em Physical Review Letters}, 104(15):153601, 2010.

\bibitem{lin2011unidirectional}
Zin Lin, Hamidreza Ramezani, Toni Eichelkraut, Tsampikos Kottos, Hui Cao, and
  Demetrios~N Christodoulides.
\newblock Unidirectional invisibility induced by p t-symmetric periodic
  structures.
\newblock {\em Physical Review Letters}, 106(21):213901, 2011.

\bibitem{lee2014heralded}
Tony~E Lee and Ching-Kit Chan.
\newblock Heralded magnetism in non-hermitian atomic systems.
\newblock {\em Physical Review X}, 4(4):041001, 2014.

\bibitem{malzard2015topologically}
Simon Malzard, Charles Poli, and Henning Schomerus.
\newblock Topologically protected defect states in open photonic systems with
  non-hermitian charge-conjugation and parity-time symmetry.
\newblock {\em Physical Review Letters}, 115(20):200402, 2015.

\bibitem{yamamoto2019theory}
Kazuki Yamamoto, Masaya Nakagawa, Kyosuke Adachi, Kazuaki Takasan, Masahito
  Ueda, and Norio Kawakami.
\newblock Theory of non-hermitian fermionic superfluidity with a complex-valued
  interaction.
\newblock {\em Physical Review Letters}, 123(12):123601, 2019.

\bibitem{li2019observation}
Jiaming Li, Andrew~K Harter, Ji~Liu, Leonardo de~Melo, Yogesh~N Joglekar, and
  Le~Luo.
\newblock Observation of parity-time symmetry breaking transitions in a
  dissipative floquet system of ultracold atoms.
\newblock {\em Nature communications}, 10(1):1--7, 2019.

\bibitem{zyuzin2018flat}
Alexander~A Zyuzin and A~Yu Zyuzin.
\newblock Flat band in disorder-driven non-hermitian weyl semimetals.
\newblock {\em Physical Review B}, 97(4):041203, 2018.

\bibitem{shen2018quantum}
Huitao Shen and Liang Fu.
\newblock Quantum oscillation from in-gap states and a non-hermitian landau
  level problem.
\newblock {\em Physical Review Letters}, 121(2):026403, 2018.

\bibitem{yoshida2018non}
Tsuneya Yoshida, Robert Peters, and Norio Kawakami.
\newblock Non-hermitian perspective of the band structure in heavy-fermion
  systems.
\newblock {\em Physical Review B}, 98(3):035141, 2018.

\bibitem{zhou2018observation}
Hengyun Zhou, Chao Peng, Yoseob Yoon, Chia~Wei Hsu, Keith~A Nelson, Liang Fu,
  John~D Joannopoulos, Marin Solja{\v{c}}i{\'c}, and Bo~Zhen.
\newblock Observation of bulk fermi arc and polarization half charge from
  paired exceptional points.
\newblock {\em Science}, 359(6379):1009--1012, 2018.

\bibitem{papaj2019nodal}
Micha{\l} Papaj, Hiroki Isobe, and Liang Fu.
\newblock Nodal arc of disordered dirac fermions and non-hermitian band theory.
\newblock {\em Physical Review B}, 99(20):201107, 2019.

\bibitem{kimura2019chiral}
Kazuhiro Kimura, Tsuneya Yoshida, and Norio Kawakami.
\newblock Chiral-symmetry protected exceptional torus in correlated nodal-line
  semimetals.
\newblock {\em Physical Review B}, 100(11):115124, 2019.

\bibitem{yao2018edge}
Shunyu Yao and Zhong Wang.
\newblock Edge states and topological invariants of non-hermitian systems.
\newblock {\em Physical Review Letters}, 121(8):086803, 2018.

\bibitem{lee2019anatomy}
Ching~Hua Lee and Ronny Thomale.
\newblock Anatomy of skin modes and topology in non-hermitian systems.
\newblock {\em Physical Review B}, 99(20):201103, 2019.

\bibitem{lee2019hybrid}
Ching~Hua Lee, Linhu Li, and Jiangbin Gong.
\newblock Hybrid higher-order skin-topological modes in nonreciprocal systems.
\newblock {\em Physical Review Letters}, 123(1):016805, 2019.

\bibitem{longhi2019probing}
Stefano Longhi.
\newblock Probing non-hermitian skin effect and non-bloch phase transitions.
\newblock {\em Physical Review Research}, 1(2):023013, 2019.

\bibitem{okuma2020topological}
Nobuyuki Okuma, Kohei Kawabata, Ken Shiozaki, and Masatoshi Sato.
\newblock Topological origin of non-hermitian skin effects.
\newblock {\em Physical Review Letters}, 124(8):086801, 2020.

\bibitem{yi2020non}
Yifei Yi and Zhesen Yang.
\newblock Non-hermitian skin modes induced by on-site dissipations and chiral
  tunneling effect.
\newblock {\em Physical Review Letters}, 125(18):186802, 2020.

\bibitem{kawabata2020higher}
Kohei Kawabata, Masatoshi Sato, and Ken Shiozaki.
\newblock Higher-order non-hermitian skin effect.
\newblock {\em Physical Review B}, 102(20):205118, 2020.

\bibitem{liu2020helical}
Chun-Hui Liu, Kai Zhang, Zhesen Yang, and Shu Chen.
\newblock Helical damping and dynamical critical skin effect in open quantum
  systems.
\newblock {\em Physical Review Research}, 2(4):043167, 2020.

\bibitem{fu2021non}
Yongxu Fu, Jihan Hu, and Shaolong Wan.
\newblock Non-hermitian second-order skin and topological modes.
\newblock {\em Physical Review B}, 103(4):045420, 2021.

\bibitem{okuma2021non}
Nobuyuki Okuma and Masatoshi Sato.
\newblock Non-hermitian skin effects in hermitian correlated or disordered
  systems: Quantities sensitive or insensitive to boundary effects and
  pseudo-quantum-number.
\newblock {\em Physical Review Letters}, 126(17):176601, 2021.

\bibitem{sun2021geometric}
Xiao-Qi Sun, Penghao Zhu, and Taylor~L Hughes.
\newblock Geometric response and disclination-induced skin effects in
  non-hermitian systems.
\newblock {\em arXiv preprint arXiv:2102.05667}, 2021.

\bibitem{claes2021skin}
Jahan Claes and Taylor~L Hughes.
\newblock Skin effect and winding number in disordered non-hermitian systems.
\newblock {\em Physical Review B}, 103(14):L140201, 2021.

\bibitem{guo2021exact}
Cui-Xian Guo, Chun-Hui Liu, Xiao-Ming Zhao, Yanxia Liu, and Shu Chen.
\newblock Exact solution of non-hermitian systems with generalized boundary
  conditions: size-dependent boundary effect and fragility of skin effect.
\newblock {\em arXiv preprint arXiv:2102.03781}, 2021.

\bibitem{kunst2018biorthogonal}
Flore~K Kunst, Elisabet Edvardsson, Jan~Carl Budich, and Emil~J Bergholtz.
\newblock Biorthogonal bulk-boundary correspondence in non-hermitian systems.
\newblock {\em Physical Review Letters}, 121(2):026808, 2018.

\bibitem{herviou2019defining}
Loic Herviou, Jens~H Bardarson, and Nicolas Regnault.
\newblock Defining a bulk-edge correspondence for non-hermitian hamiltonians
  via singular-value decomposition.
\newblock {\em Physical Review A}, 99(5):052118, 2019.

\bibitem{imura2019generalized}
Ken-Ichiro Imura and Yositake Takane.
\newblock Generalized bulk-edge correspondence for non-hermitian topological
  systems.
\newblock {\em Physical Review B}, 100(16):165430, 2019.

\bibitem{lee2019topological}
Jong~Yeon Lee, Junyeong Ahn, Hengyun Zhou, and Ashvin Vishwanath.
\newblock Topological correspondence between hermitian and non-hermitian
  systems: Anomalous dynamics.
\newblock {\em Physical Review Letters}, 123(20):206404, 2019.

\bibitem{wang2019non}
Huaiqiang Wang, Jiawei Ruan, and Haijun Zhang.
\newblock Non-hermitian nodal-line semimetals with an anomalous bulk-boundary
  correspondence.
\newblock {\em Physical Review B}, 99(7):075130, 2019.

\bibitem{jin2019bulk}
L~Jin and Z~Song.
\newblock Bulk-boundary correspondence in a non-hermitian system in one
  dimension with chiral inversion symmetry.
\newblock {\em Physical Review B}, 99(8):081103, 2019.

\bibitem{yang2020non}
Zhesen Yang, Kai Zhang, Chen Fang, and Jiangping Hu.
\newblock Non-hermitian bulk-boundary correspondence and auxiliary generalized
  brillouin zone theory.
\newblock {\em Physical Review Letters}, 125(22):226402, 2020.

\bibitem{zhang2020correspondence}
Kai Zhang, Zhesen Yang, and Chen Fang.
\newblock Correspondence between winding numbers and skin modes in
  non-hermitian systems.
\newblock {\em Physical Review Letters}, 125(12):126402, 2020.

\bibitem{zhang2020bulk}
Zhicheng Zhang, Zhesen Yang, and Jiangping Hu.
\newblock Bulk-boundary correspondence in non-hermitian hopf-link exceptional
  line semimetals.
\newblock {\em Physical Review B}, 102(4):045412, 2020.

\bibitem{wang2020defective}
Xiao-Ran Wang, Cui-Xian Guo, and Su-Peng Kou.
\newblock Defective edge states and number-anomalous bulk-boundary
  correspondence in non-hermitian topological systems.
\newblock {\em Physical Review B}, 101(12):121116, 2020.

\bibitem{xue2021simple}
Wen-Tan Xue, Ming-Rui Li, Yu-Min Hu, Fei Song, and Zhong Wang.
\newblock Simple formulas of directional amplification from non-bloch band
  theory.
\newblock {\em Physical Review B}, 103(24):L241408, 2021.

\bibitem{cao2021non}
Yang Cao, Yang Li, and Xiaosen Yang.
\newblock Non-hermitian bulk-boundary correspondence in a periodically driven
  system.
\newblock {\em Physical Review B}, 103(7):075126, 2021.

\bibitem{yokomizo2019non}
Kazuki Yokomizo and Shuichi Murakami.
\newblock Non-bloch band theory of non-hermitian systems.
\newblock {\em Physical Review Letters}, 123(6):066404, 2019.

\bibitem{kawabata2020non}
Kohei Kawabata, Nobuyuki Okuma, and Masatoshi Sato.
\newblock Non-bloch band theory of non-hermitian hamiltonians in the symplectic
  class.
\newblock {\em Physical Review B}, 101(19):195147, 2020.

\bibitem{yokomizo2021non}
Kazuki Yokomizo and Shuichi Murakami.
\newblock Non-bloch band theory in bosonic bogoliubov--de gennes systems.
\newblock {\em Physical Review B}, 103(16):165123, 2021.

\bibitem{song2019non}
Fei Song, Shunyu Yao, and Zhong Wang.
\newblock Non-hermitian topological invariants in real space.
\newblock {\em Physical Review Letters}, 123(24):246801, 2019.

\bibitem{deng2019non}
Tian-Shu Deng and Wei Yi.
\newblock Non-bloch topological invariants in a non-hermitian domain wall
  system.
\newblock {\em Physical Review B}, 100(3):035102, 2019.

\bibitem{gong2018topological}
Zongping Gong, Yuto Ashida, Kohei Kawabata, Kazuaki Takasan, Sho Higashikawa,
  and Masahito Ueda.
\newblock Topological phases of non-hermitian systems.
\newblock {\em Physical Review X}, 8(3):031079, 2018.

\bibitem{yin2018geometrical}
Chuanhao Yin, Hui Jiang, Linhu Li, Rong L{\"u}, and Shu Chen.
\newblock Geometrical meaning of winding number and its characterization of
  topological phases in one-dimensional chiral non-hermitian systems.
\newblock {\em Physical Review A}, 97(5):052115, 2018.

\bibitem{jiang2018topological}
Hui Jiang, Chao Yang, and Shu Chen.
\newblock Topological invariants and phase diagrams for one-dimensional
  two-band non-hermitian systems without chiral symmetry.
\newblock {\em Physical Review A}, 98(5):052116, 2018.

\bibitem{kawabata2019symmetry}
Kohei Kawabata, Ken Shiozaki, Masahito Ueda, and Masatoshi Sato.
\newblock Symmetry and topology in non-hermitian physics.
\newblock {\em Physical Review X}, 9(4):041015, 2019.

\bibitem{kawabata2019classification}
Kohei Kawabata, Takumi Bessho, and Masatoshi Sato.
\newblock Classification of exceptional points and non-hermitian topological
  semimetals.
\newblock {\em Physical Review Letters}, 123(6):066405, 2019.

\bibitem{zhou2019periodic}
Hengyun Zhou and Jong~Yeon Lee.
\newblock Periodic table for topological bands with non-hermitian symmetries.
\newblock {\em Physical Review B}, 99(23):235112, 2019.

\bibitem{yokomizo2020topological}
Kazuki Yokomizo and Shuichi Murakami.
\newblock Topological semimetal phase with exceptional points in
  one-dimensional non-hermitian systems.
\newblock {\em Physical Review Research}, 2(4):043045, 2020.

\bibitem{Supp1}
See Supplemental Material at [URL will be inserted by publisher] for (i)
  numberical method of calculating the OBC spectrum and GBZ, (ii) Non-Hermitian
  systems with their GBZ being a perfect circle, (iii) proof of correspondence
  between winding number and OBC spectrum in anomalous time-reversal symmetry
  systems, (iv) reality of the 3-bifurcation states.

\bibitem{Supp2}
We have provided some models as examples of this exceptional case in
  Supplemental Material [77].

\bibitem{Supp3}
For a simple model, we have a detailed proof of the existence of the
  3-bifuraction states on the open-boundary condition spectrum, see
  Supplemental Material [77].

\bibitem{Supp4}
The proof is similar to the previous one in proving the theorem, see
  Supplemental Material [77] for detail.

\end{thebibliography}


\begin{thebibliography}{0}
\expandafter\ifx\csname natexlab\endcsname\relax\def\natexlab#1{#1}\fi
\expandafter\ifx\csname bibnamefont\endcsname\relax
  \def\bibnamefont#1{#1}\fi
\expandafter\ifx\csname bibfnamefont\endcsname\relax
  \def\bibfnamefont#1{#1}\fi
\expandafter\ifx\csname citenamefont\endcsname\relax
  \def\citenamefont#1{#1}\fi
\expandafter\ifx\csname url\endcsname\relax
  \def\url#1{\texttt{#1}}\fi
\expandafter\ifx\csname urlprefix\endcsname\relax\def\urlprefix{URL }\fi
\providecommand{\bibinfo}[2]{#2}
\providecommand{\eprint}[2][]{\url{#2}}

\end{thebibliography}

\end{document}


\title{Supplemental Material for ``Connections between the Open-boundary Spectrum and Generalized Brillouin Zone in Non-Hermitian Systems''}
\author{Deguang Wu}
\affiliation{National Laboratory of Solid State Microstructures, Department of Physics, Nanjing University, Nanjing 210093, China}
\author{Jiao Xie}
\affiliation{National Laboratory of Solid State Microstructures, Department of Physics, Nanjing University, Nanjing 210093, China}
\author{Yao Zhou}
\affiliation{National Laboratory of Solid State Microstructures, Department of Physics, Nanjing University, Nanjing 210093, China}
\author{Jin An}
\email{anjin@nju.edu.cn}
\affiliation{National Laboratory of Solid State Microstructures, Department of Physics, Nanjing University, Nanjing 210093, China}
\affiliation{Collaborative Innovation Center of Advanced Microstructures, Nanjing University, Nanjing 210093, China}

\date{\today}

\maketitle

\subsection{Numerical method of calculating the OBC Spectrum and the GBZ}
We first review our new numerical method that can accurately obtain the OBC spectrum and the GBZ for non-Hermitian systems.
As mentioned in the main article, the map of $|\beta|=\rho$ onto the complex $E$-plane by $E_{\alpha}(\beta)$ are $N$ curves $E_{\alpha}(|\beta|=\rho)(\alpha=1,2,\ldots,N)$, which form some closed loops.
There exist some crossing points, which can be self-intersection ones between the same loop or intersection ones between different loops.
The method of calculating the OBC spectrum and the GBZ based on our theorem can be summarized as follows:

(1). Find the crossing points of these closed loops.

(2). Pick out these crossing points which are on the boundary between two areas with winding number $w_{\rho}=+1$ and $w_{\rho}=-1$ respectively.

(3). Increase $\rho$ gradually from zero to infinity, then one can obtain the whole OBC spectrum and the GBZ.
\begin{figure}
    \centering
    \includegraphics[width=0.48\textwidth]{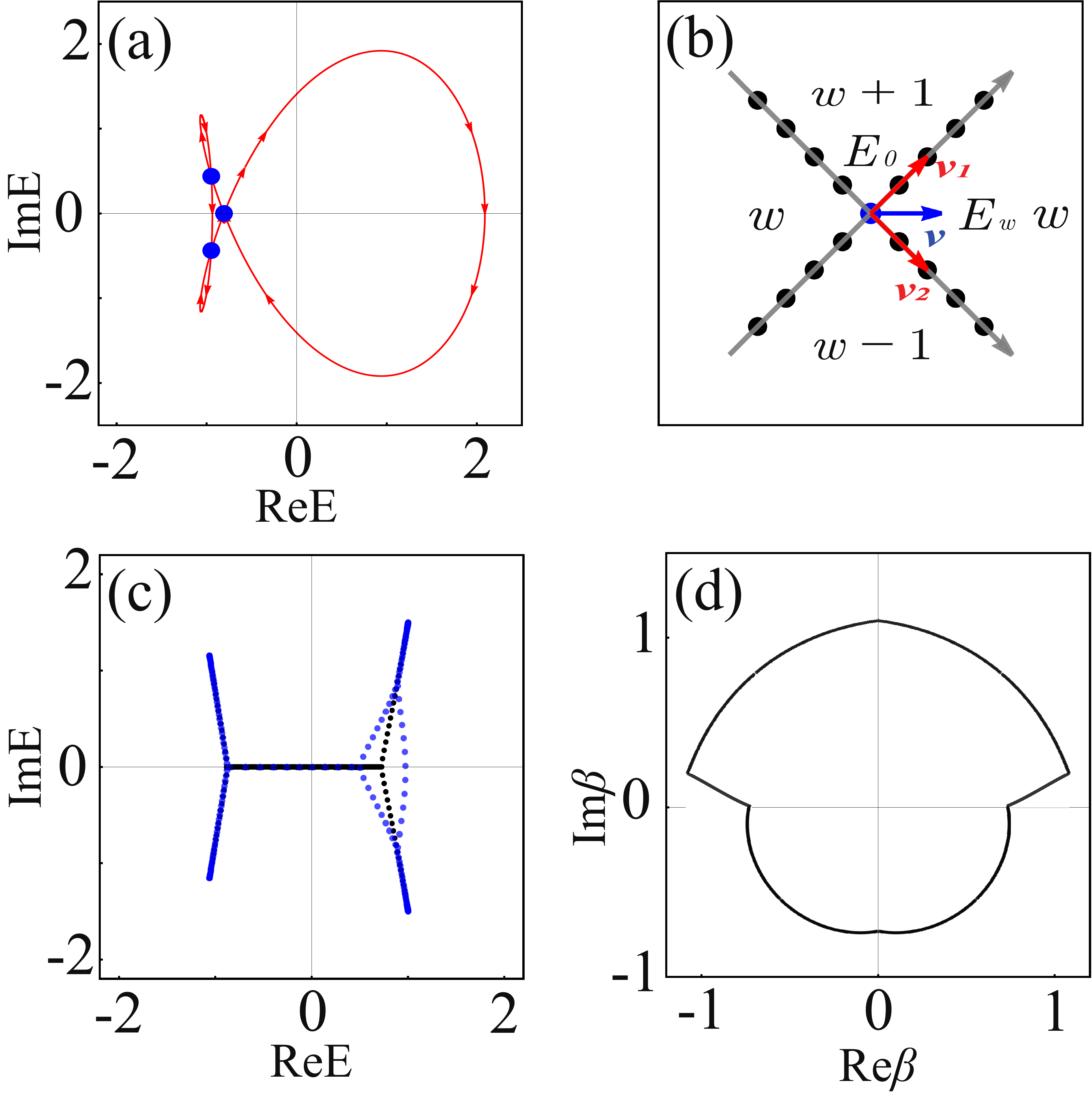}
    \caption{Illustration of our numerical method of calculating the OBC spectrum and the GBZ.
    (a) A PBC spectrum with several self-crossing points (blue dots).
    (b) One of the crossing points and the winding numbers distribution adjacent to it.
    (c) Comparison between our numerical method (black points) and the numerical diagonalization method (blue points) with lattice size and digit precision chosen as $N=200$, $P=500$.
    (d) GBZ calculated from our numerical method.}
    \label{fig1}
\end{figure}

To show our method, we start with a single-band model with Hamiltonian $H(\beta)=-\frac{1}{\beta^2}+\frac{\mathrm{i}}{\beta}-\frac{\mathrm{i}}{2}\beta+\frac{1}{2}\beta^{2}$ as an example.
The PBC spectrum $E(|\beta|=\rho)$ forms a closed loop, for which there may exist some self-intersection points.
First of all, we seek out the crossing points as shown in Fig1.(a).
Then among them we will pick out those points with their two adjacent areas possessing winding numbers $-1/+1$, which belong to the OBC spectrum.

Now we consider one of the crossing points of the PBC spectrum, which we label as $E_{0}$, the winding number distribution in the areas adjacent to $E_{0}$ can only be $(w+1)/w/(w-1)/w$, as shown in Fig1.(b). If $w=0$, then $E_{0}$ is on the OBC spectrum.

In order to check whether $w=0$, we first select the points belonging to the PBC spectrum that satisfy the condition of $|E(|\beta|=\rho)-E_{0}|<\delta$.
Here $\delta$ is a small quantity artificially set, which needs to be adjusted according to the specific situation.
In Fig1.(b), we plot the crossing point $E_{0}$ of the PBC spectrum and several points near $E_{0}$.
Then in the direction of the PBC spectrum, we determine two vectors $\boldsymbol{v_{1}}$,$\boldsymbol{v_{2}}$ along the two arcs of the spectrum starting from $E_{0}$. We can define a vector $\boldsymbol{v}=\frac{1}{2}(\boldsymbol{v_{1}}+\boldsymbol{v_{2}})$, which points to the region where the winding number is $w$. Use $\boldsymbol{v}$ to shift $E_{0}$ to a position, named $E_{w}$, at which, we calculate the winding number. If $w=0$, reserve $E_{0}$ as a state belonging to the OBC spectrum. To calculate the value of $w$, we use the definition in the main article $w=w_{\rho}(E_{w})=N_{\rho}(E_{w})-p$, with $N_{\rho}(E)$ is the number of $\beta_{i}(E)$ inside the circle $|\beta|=\rho$.
As the value of $\rho$ goes from zero to infinity, we retain the points that meet the conditions referred to above, so the OBC spectrum can be completely determined as shown in Fig1.(c).
Accordingly, the GBZ (See Fig1.(d)) can also be determined.

We further point out that the magnitude of $\boldsymbol{v}$ should be selected appropriately to ensure that $E_{w}$ should be not only sufficiently separated from $E_{0}$, but also be within the region possessing the winding number $w$.

The above method can be applied to both single-band and multi-band Hamiltonians to efficiently get more focused results, compared to the conventional numerical diagonalization method, which is often sensitive to the matrix dimension and calculation accuracy, as shown in Fig.1(c).
In the calculations, there are $20,000$ points on the OBC spectrum that can be determined efficiently by our method.
In contrast, this needs to diagonalize the Hamiltonian $H_{N \times N}$ with $N=20,000$ by using numerical diagonalization method, which become incorrect and time-consuming.

\subsection{Non-Hermitian systems with their GBZ being a perfect circle}
An exception to our theorem is when the GBZ is a perfect circle $|\beta|=r$. This situation occurs when the PBC spectrum $E(|\beta|=r)$ coincides with the whole OBC spectrum, and the other PBC spectra $E(|\beta|=\rho)$ with $\rho\neq r$ give no contributions to the OBC spectrum.
Here, we give some models to demonstrate this exceptional case.

Firstly, we consider the following two-band model,
\begin{equation}
    H(\beta)=(\beta+\frac{1}{\beta})\sigma_{1}+\{a(\beta-\frac{1}{\beta})+c\}\sigma_{3}.
\end{equation}
Its energy bands can be obtained as
\begin{equation}
    E=\pm\sqrt{1+a^2}(\beta-\frac{1}{\beta})\pm2\mathrm{i}a,
\end{equation}
if $c$ is chosen to be $c=2\mathrm{i}\sqrt{1+a^2}$, where $a\in\mathbb{C}$.
Since $\{\sigma_{2},H\}=0$, this model has chiral symmetry, indicating the OBC spectrum should be centrosymmetric. As indicated in Fig.2(a), the OBC spectrum is composed of two line segments, which coincides on the $E$-plane with the PBC spectrum $E(|\beta|=1)$, while the GBZ is the unit circle $|\beta|=1$.

We can construct a more generic two-band model whose Hamiltonian reads
\begin{equation}
   H(\beta)=\sum_{j=1}^{3}d_{j}(\beta)\sigma_{j},
\end{equation}
where $d_{j}=a_{j}\beta+b_{j}\frac{1}{\beta}+c_{j}$, and $a_{j}$, $b_{j}$, $c_{j}\in\mathbb{C}$ with $j=1,2,3$.
These parameters can be combined into three vectors $\boldsymbol{a}=(a_{1},a_{2},a_{3})$, $\boldsymbol{b}=(b_{1},b_{2},b_{3})$, $\boldsymbol{c}=(c_{1},c_{2},c_{3})$.
We find that if $\boldsymbol{c}$ is chosen to be
\begin{equation}
    \boldsymbol{c}=\frac{\sqrt{2}\boldsymbol{a}\times\boldsymbol{b}}{\sqrt{\sqrt{\boldsymbol{a}^2}\sqrt{\boldsymbol{b}^2}+\boldsymbol{a}\cdot\boldsymbol{b}}},
\end{equation}
then the two energy bands can be obtained as
\begin{equation}
E=\pm(\sqrt{\boldsymbol{a}^{2}}\beta+\sqrt{\boldsymbol{b}^{2}}\frac{1}{\beta}),
\end{equation}
where $\sqrt{\boldsymbol{a}^2}$ and $\sqrt{\boldsymbol{b}^2}$ can be chosen to be any one of the two square roots of $\boldsymbol{a}^2$ and $\boldsymbol{b}^2$, respectively. While the GBZ is still a perfect circle $|\beta|=r$, the segment-like OBC spectrum coincides with the PBC spectrum $E(|\beta|=r)$, as illustrated in Fig.2(b).

We now provide another kind of two-band models possessing the unit circle as their GBZs. These models have the following Hamiltonian:
\begin{equation}
    H(\beta) = \sum_{j=1}^{3}({d}_{0j} \mathbb{I}+\boldsymbol{d}_{j}\cdot\boldsymbol{\sigma})(\beta^{j}+\beta^{-j}),
\end{equation}
where $\boldsymbol{\sigma}=({\sigma_{1},\sigma_{2},\sigma_{3}})$, $\boldsymbol{d}_{j}=(d_{j}^{1},d_{j}^{2},d_{j}^{3})(j=1,2,3)$, and $d_{0j},d_{j}^{k}\in\mathbb{C}$ with $j,k=1,2,3$. Obviously, $H(\beta)=H(\beta^{-1})$, indicating that the system has inversion symmetry.
The OBC spectrum of this model could be two complicated arcs, which always coincides exactly with the PBC spectrum $E(|\beta|=1)$, as shown in Fig.2(c-d), while the GBZ is still the unit circle. Particularly, when all $d_{0j}=0$, the OBC spectrum has central symmetry with respect to the origin $E=0$(See Fig.2(d)).
\begin{figure}
    \centering
    \includegraphics[width=0.48\textwidth]{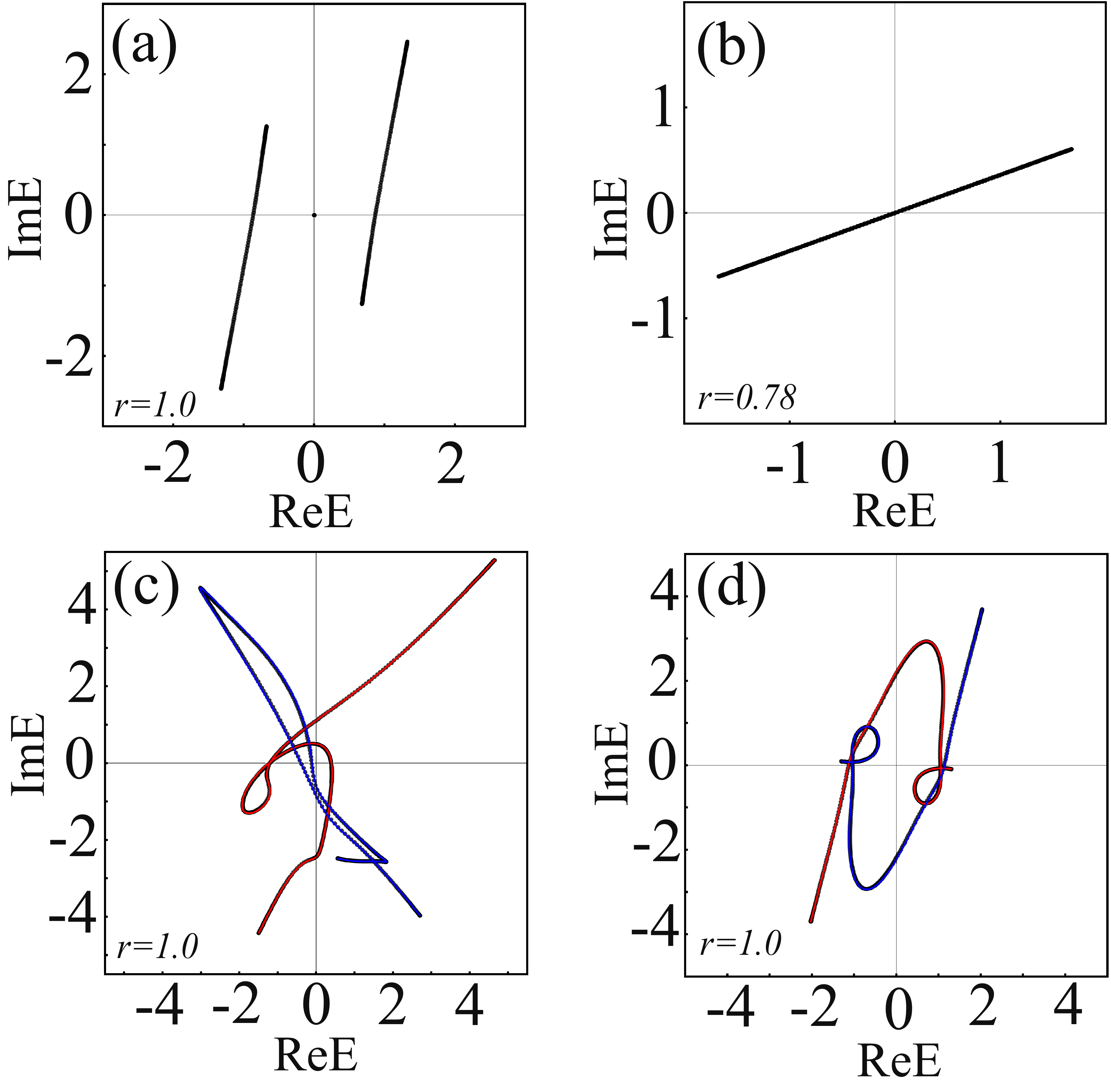}
    \caption{OBC spectra of two-band systems with their GBZ being a perfect circle $|\beta|=r$, where each OBC spectrum coincides exactly with its PBC spectrum $E(|\beta|=r)$. (a) Model given by Eq.(1) with the parameters chosen as $a=-0.3+0.5\mathrm{i}$.
    (b) Model given by Eq.(3) with the parameters $\boldsymbol{a}=(0.8+0.8\mathrm{i},0.5\mathrm{i},-0.7)$, $\boldsymbol{b}=(-0.6,0.4,0.2\mathrm{i})$.
    (c-d) Model given by Eq.(6) with the nonzero parameters taken as $(d_{01},\boldsymbol{d}_{1})=(0.5+0.4\mathrm{i}, \mathrm{i}, 0.2-0.1\mathrm{i}, 0.3+0.3\mathrm{i})$, $(d_{02},\boldsymbol{d}_{2})=(0.8-0.7\mathrm{i}, 0.7\mathrm{i}, 0.3\mathrm{i}, 0.45\mathrm{i})$, $(d_{03},\boldsymbol{d}_{3})=(\mathrm{i}, 0.73, 0.5\mathrm{i}, 0.5)$ for (c),
    and $\boldsymbol{d}_{1}=(0.2,0.6,0.1)$, $\boldsymbol{d}_{2}=(0.6+\mathrm{i}, 0.4\mathrm{i}, 0.3\mathrm{i})$, $\boldsymbol{d}_{3}=(0.7\mathrm{i}, 0.3\mathrm{i}, 0.2)$ for (d).}
    \label{fig2}
\end{figure}

\subsection{Proof of the correspondence between winding number and the OBC spectrum in systems with anomalous time-reversal symmetry}
Here, we give a proof of the central conclusion generalized to systems in the symplectic class in the main article.
\begin{figure}
    \centering
    \includegraphics[width=0.48\textwidth]{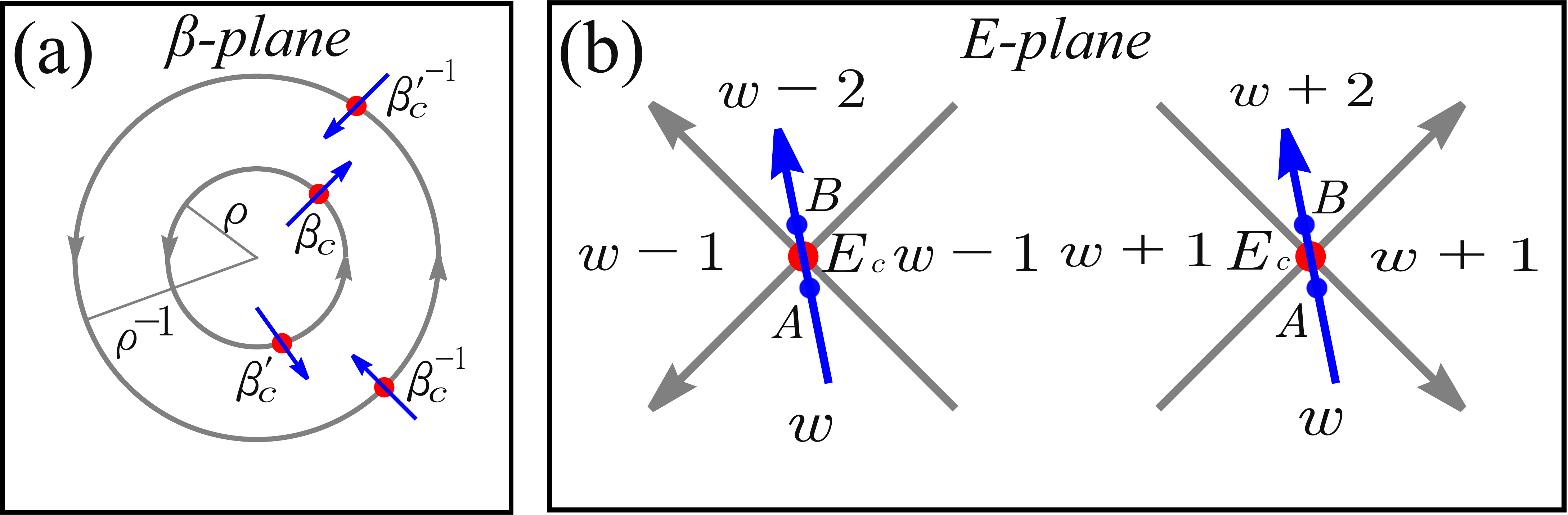}
    \caption{Schematic depiction of the proof for systems in the symplectic class.
    (a) $|\beta|=\rho$ and $|\beta|=\rho^{-1}$ with $\rho<1$, (b) One of the crossing points $E_c$ of the corresponding periodic-boundary spectrum $E(|\beta|=\rho)$ or $E(|\beta|=\rho^{-1})$, where the two loops coincide with each other but point in opposite directions due to the anomalous time-reversal symmetry, and the four areas adjacent to $E_c$ have been marked by their winding numbers. The spectra on the $E$-plane are the images of the two circles on the $\beta$-plane, under the mapping $E(\beta)$. The four solid dots $\beta_c$, $\beta_c^{'}$, $\beta_c^{-1}$ and $\beta_c^{'-1}$ on the two circles in (a) are the preimages of $E_c$, and accordingly, the four small arcs passing through the four points correspond to the trajectory from $A$ to $B$ in (b).}
    \label{fig3}
\end{figure}
As shown in Fig.3(b), the pair of loops of PBC spectrum $E(|\beta|=\rho)$ and $E(|\beta|=\rho^{-1})$ overlap completely, but point in opposite directions. Assume $\rho<1$ and let $E_{c}$ be a crossing point of the PBC spectrum $E(|\beta|=\rho)$ or $E(|\beta|=\rho^{-1})$, then there must be two different $\beta_{c}$ and $\beta_{c}^{\prime}$ on the circle $|\beta|=\rho$, while $\beta_{c}^{-1}$ and $\beta_{c}^{\prime-1}$ are on the circle $|\beta|=\rho^{-1}$ (see Fig.3(a)). Next, we prove that the prerequisite for $E_c$ being on the OBC spectrum is $w=0$.
Consider a small curve from point $A$ through point $E_c$ to point B. Due to Eq.(4) in the main article, the number of $\beta_i(E_{A/B})$ inside the circle $|\beta|=\rho$ should be $(p+w)/(p+w-2)$, and the number of $\beta_i(E_{A/B})$ inside the circle $|\beta|=\rho^{-1}$ should be $(p+w)/(p+w+2)$ correspondingly. If $A$ and $B$ are sufficiently close to $E_c$, among the $2p$ image curves of the map $\beta_i(E)$ of the trajectory $E_A \to E_c \to E_B$, there are two possible scenarios that will happen, as shown in Fig.3(a), one is that two small outward curves would intersect the cirle $|\beta|=\rho$ at $\beta_c$ and $\beta_c^{'}$, the other is that two small inward curves would intersect the cirle $|\beta|=\rho^{-1}$ at $\beta_c^{-1}$ and $\beta_c^{'-1}$. The former corresponds to the change in winding number from $w$ to $w-2$, the latter corresponds to the change in winding number from $w$ to $w+2$. If $w=0$, then $\beta_c$ and $\beta_c^{'}$ must be $(p-1)$th and $p$th $\beta$ solutions of $E_c$, $\beta_c^{-1}$ and $\beta_c^{'-1}$ must be $(p+1)$th and $(p+2)$th $\beta$ solutions of $E_c$ simultaneously. In other words, both the equations $|\beta_{p-1}(E_c)|=|\beta_{p}(E_c)|$ and $|\beta_{p+1}(E_c)|=|\beta_{p+2}(E_c)|$ hold, which means that $E_c$ is on the OBC spectrum.
So for systems in the symplectic class, the crossing points of the PBC spectrum with its two adjacent areas possessing winding numbers $0/\pm2$ must belong to the OBC spectrum. The inverse proposition can be analogously proved by imitating the main article.

\subsection{Reality of the 3-bifurcation states}
In this section, we first give an explanation of the reality of the 3-bifurcation states, and then we further give a two-band model without symmetry and a four-band model with anomalous time-reversal symmetry as examples to exhibit the features of the 3-bifurcation and 4-bifurcation states.

One may ask a  question: whether are the 3-bifurcation states really on the OBC spectrum?
Namely, whether is a 3-bifurcation state just a limiting point approaching from the three corresponding branches of the OBC spectrum?
Here, we demonstrate by a simple model that the 3-bifurcation states do exist and actually belong to the OBC spectrum.

Let's consider a single-band model with Hamiltonian $H(\beta)=-\frac{1}{2}\beta^{-2}+\frac{\mathrm{i}}{2}\beta^{-1}-\frac{\mathrm{i}}{2}\beta+\frac{1}{2}\beta^2$, whose characteristic equation $E-H(\beta)=0$ can be equivalently written as
\begin{equation}
    \beta^{4}-\mathrm{i}\beta^{3}-2E\beta^{2}+\mathrm{i}\beta-1=0.
\end{equation}
The OBC spectrum and GBZ of this model are shown in Fig.4 as a comparison.
\begin{figure}
    \centering
    \includegraphics[width=0.48\textwidth]{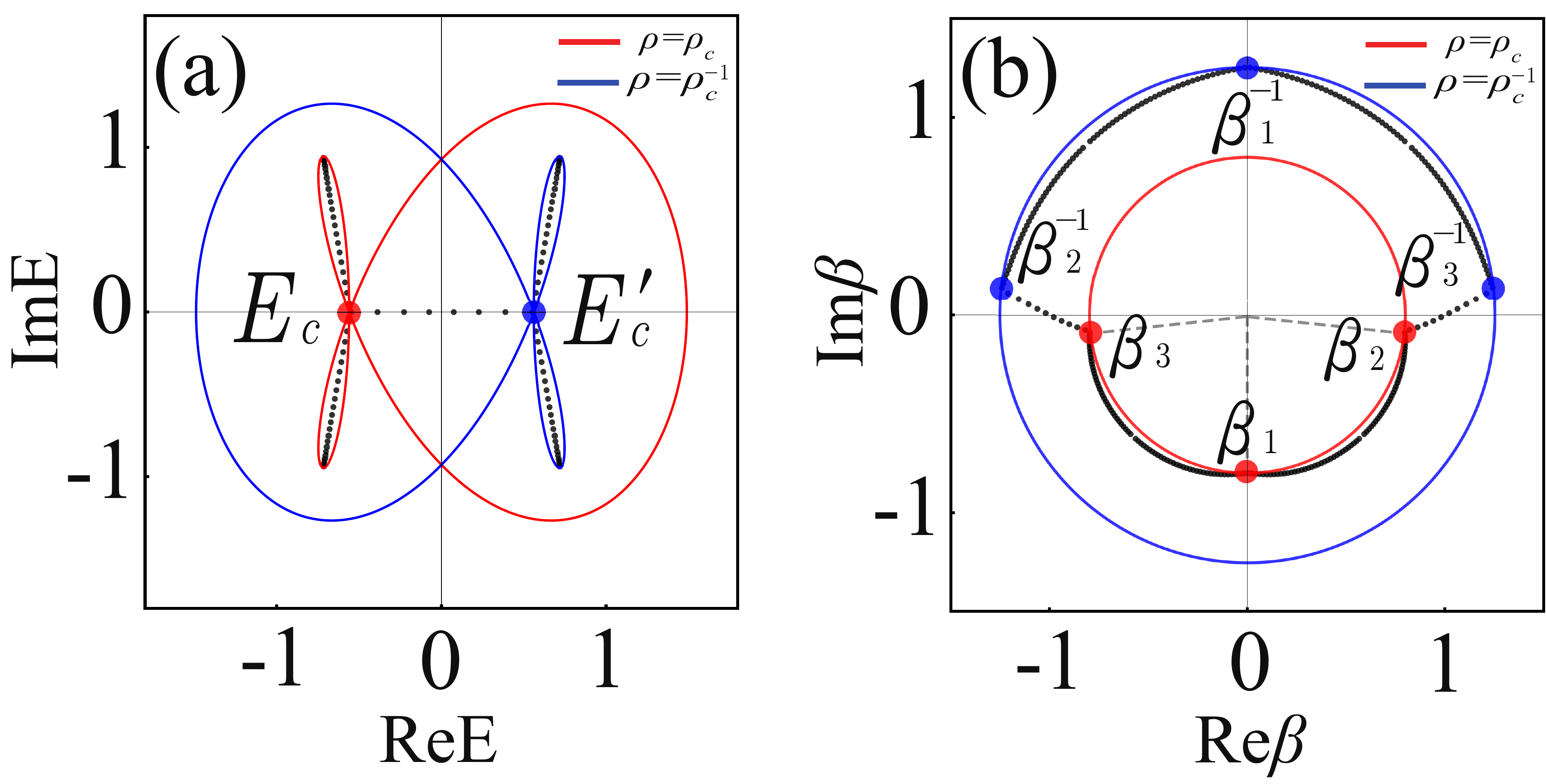}
    \caption{(a) OBC spectrum(solid dots) together with the PBC spectra(solid lines) with $\rho=\rho_{c}$ , $\rho_{c}^{-1}$ passing through the 3-bifurcation states, (b) GBZ of the model corresponding to Eq.(7). The red(blue) circle $|\beta|=\rho_{c}$ ($|\beta|=\rho_{c}^{-1}$) intersects the GBZ at three points $\beta_{1}$, $\beta_{2}$, $\beta_{3}$($\beta_{1}^{-1}$, $\beta_{2}^{-1}$, $\beta_{3}^{-1}$) which are the counterpart of the GBZ for the 3-bifurcation state $E_{c}(E^{'}_{c})$ on the OBC spectrum. Here $\rho_{c}=0.7987$.}
    \label{fig4}
\end{figure}
According to the relations between roots and coefficients of the quartic equation, the four roots $\beta_{i}$($i=1,2,3,4$) satisfy
\begin{gather}
    \beta_{1}\beta_{2}\beta_{3}\beta_{4}=-1,\\
    \beta_{1}+\beta_{2}+\beta_{3}+\beta_{4}=\mathrm{i},\\
    \beta_{1}^{-1}+\beta_{2}^{-1}+\beta_{3}^{-1}+\beta_{4}^{-1}=\mathrm{i},\\
    \beta_{1}\beta_{2}+\beta_{1}\beta_{3}+\beta_{1}\beta_{4}+\beta_{2}\beta_{3}+\beta_{2}\beta_{4}+\beta_{3}\beta_{4}=-2E.
\end{gather}

In order to find the 3-bifurcation states, we seek the following kind of solutions, which naturally satisfy Eq.(8),
\begin{equation}
    \begin{cases}
        \beta_{1}=-\mathrm{i}\rho_{c},\\
        \beta_{2}=-\mathrm{i}\rho_{c}e^{\mathrm{i}\theta_{0}},\\
        \beta_{3}=-\mathrm{i}\rho_{c}e^{-\mathrm{i}\theta_{0}},\\
        \beta_{4}=\mathrm{i}\rho_{c}^{-3},
    \end{cases}
\end{equation}
where $\rho_{c}$ is a positive real number and $\theta_{0}$ is a real phase angle. The two parameters $\rho_{c}$ and $\theta_{0}$ must be consistent with Eq.(9) and Eq.(10), which indicates,
\begin{gather}
    (1+2\cos\theta_{0})\rho_{c}^4+\rho_{c}^3=1,\\
    \rho_{c}^4+\rho_{c}=1+2\cos\theta_{0}.
\end{gather}
This implies
\begin{gather}
    \rho_{c}^8+\rho_{c}^5+\rho_{c}^3=1,\\
    \cos\theta_{0}=\frac{1}{2}(\rho_{c}^4+\rho_{c}-1).
\end{gather}
Obviously, Eq.(15) has a unique solution since the leftside of the equation is an increasing function of $\rho_{c}$ when $\rho_{c}>0$. Then $\cos\theta_{0}$ is also determined uniquely due to Eq.(16). We find $\rho_{c}=0.7987$, $\theta_{0}=\pm0.4672\pi$. The energy $E$ for this solution, now denoted by $E_{c}$, can be derived from Eq.(11) to be
\begin{equation}
    E_{c}=\frac{1}{2}(\rho_{c}^3-\rho_{c}^{-1})(\rho_{c}^3+1).
\end{equation}
Substitution the value of $\rho_{c}$ into the above expression leads to $E_{c}=-0.5605$. Since this solution obeys: $|\beta_{1}|=|\beta_{2}|=|\beta_{3}|<|\beta_{4}|$, we have $\beta_{i}(E_c)=\beta_{i}(i=1,2,3,4)$. So $\beta_{1}$, $\beta_{2}$ and $\beta_{3}$ are on the GBZ and $E_{c}$ is the 3-bifurcation state we are seeking. Another 3-bifurcation state can be found as follows. The solutions to Eq.(7) always come in pairs. It can be easily checked that if $(\beta_{1},\beta_{2},\beta_{3},\beta_{4})$ is a solution for energy $E$, then $(\beta_{1}^{-1},\beta_{2}^{-1},\beta_{3}^{-1},\beta_{4}^{-1})$ must be a solution for $-E$. Therefore, the energy $E'_{c}$ of another 3-bifurcation state must be $E'_{c}=-E_{c}$ and $\beta_{i}(E'_{c})=\beta_{5-i}^{-1}(i=1,2,3,4)$. As shown in Fig.4, these analytical results agree with the numerical calculations. Therefore, we assert that the 3-bifurcation states actually belong to the OBC spectrum.
\begin{figure}
    \centering
    \includegraphics[width=0.48\textwidth]{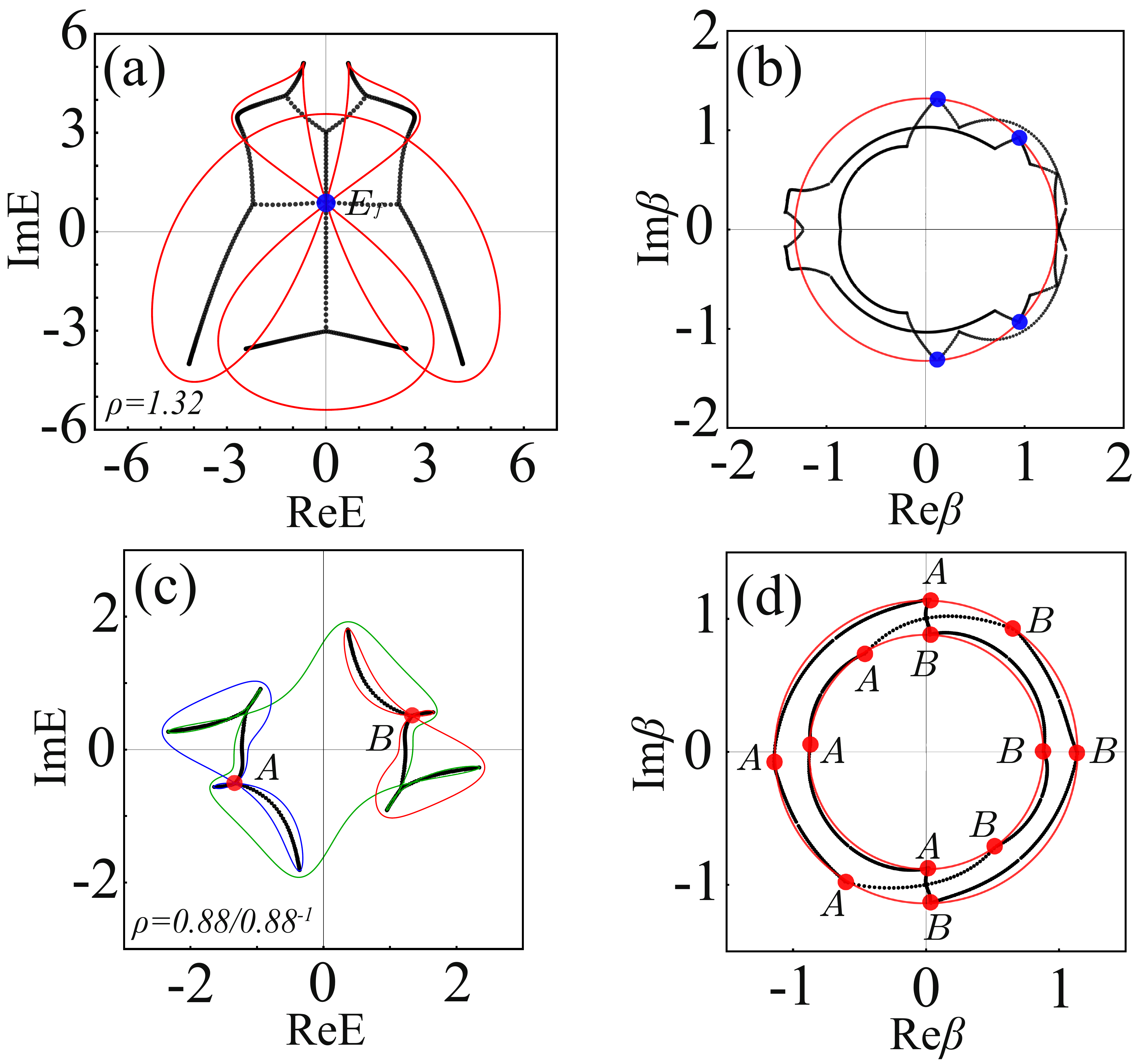}
    \caption{N-bifurcation states on the OBC spectrum and their counterparts on the GBZ.
    (a)4-bifurcation state and (c)3-bifurcation states on the OBC spectrum, together with the PBC spectrum(spectra) passing through it(them), where the former belongs to a two-band system without any symmetry while the latter belongs to a four-band system in the symplectic symmetry class.
    (b) and (d): GBZ, where the solid dots are the corresponding $\beta$s for the 4(3)-bifurcation states.}
    \label{fig5}
\end{figure}

As explained in the main article, the 3(4)-bifurcation states which belong to the OBC spectrum of a system without any symmetry can be described by the following equation $|\beta_{p-1}|=|\beta_{p}|=|\beta_{p+1}|$ or $|\beta_{p}|=|\beta_{p+1}|=|\beta_{p+2}|$ ($|\beta_{p-2}|=|\beta_{p-1}|=|\beta_{p}|=|\beta_{p+1}|$, $|\beta_{p-1}|=|\beta_{p}|=|\beta_{p+1}|=|\beta_{p+2}|$, or $|\beta_{p}|=|\beta_{p+1}|=|\beta_{p+2}|=|\beta_{p+3}|$).
Nevertheless, for systems with anomalous time-reversal symmetry, the 3(4)-bifurcation states obey $|\beta_{p-2}|=|\beta_{p-1}|=|\beta_{p}|$ or $|\beta_{p+1}|=|\beta_{p+2}|=|\beta_{p+3}|$ ($|\beta_{p-3}|=|\beta_{p-2}|=|\beta_{p-1}|=|\beta_{p}|$ or $|\beta_{p+1}|=|\beta_{p+2}|=|\beta_{p+3}|=|\beta_{p+4}|$). In Fig.(5), some more examples with 3-bifurcation states and 4-bifurcation states are demonstrated. Here the two-band model is without any symmetry, which can be described in terms of notations in Eq.(2) in the main article.
In this two-band model, $p=q=4$, and the nonzero parameters are chosen as $T_{-2}=0.5\mathrm{i} \mathbb{I}+1.5\mathrm{i} \sigma_{3}$, $T_{-1}=T_{2}=2\mathrm{i} \mathbb{I}$, $T_{0}=\sigma_{1}$, $T_{1}=3\mathrm{i} \mathbb{I}$. The clothes-like OBC spectrum possesses a 4-bifurcation state, where a PBC spectrum with a definite $\rho$ would be passing through it four times. For this 4-bifurcation state, there are four $\beta_{i}(E)$ solutions with the same absolute value located on the GBZ, namely $|\beta_{4}|=|\beta_{5}|=|\beta_{6}|=|\beta_{7}|$. These behaviors are exhibited in Fig.(5)(a-b). The other model is a minimal four-band model given below,
\begin{align}
        H(\beta)&=\Delta\gamma_{2}\nonumber\\
        &+ (a_{1}\gamma_{1}+a_{2}\gamma_{2}+a_{3}\gamma_{3}+a_{4}\gamma_{4}+a_{12} \Gamma_{12})\beta\nonumber\\
        &+ (-a_{1}\gamma_{1}+a_{2}\gamma_{2}-a_{3}\gamma_{3}-a_{4}\gamma_{4}+a_{12} \Gamma_{12})\frac{1}{\beta}\nonumber\\
        &+ (d\gamma_{1}\beta^{2}-d\gamma_{1}\frac{1}{\beta^{2}}),
\end{align}
where $\gamma_{\mu}(\mu=1,2,3,4,5)$ are the $4\times4$ gamma matrices satisfying Clifford algebra, namely, $\{\gamma_{\mu}, \gamma_{\nu}\}=2\delta_{\mu\nu}$, and $\Gamma_{12}=\mathrm{i}\gamma_{1}\gamma_{2}$. Here we choose $\gamma_{1,2,3}=\sigma_{1}\bigotimes \sigma_{1,2,3}$, $\gamma_{4}=\sigma_{3}\bigotimes \mathbb{I}$, $\gamma_{5}=\sigma_{2}\bigotimes \mathbb{I}=\gamma_{1}\gamma_{2}\gamma_{3}\gamma_{4}$.
It can be easily checked that the non-Bloch Hamiltonian obeys $\gamma_{5}{H}^{t}(\beta^{-1})\gamma_{5}^{-1}=H(\beta)$, so the system described by this Hamiltonian indeed belongs to the symplectic symmetry class. When these nonzero parameters are chosen as $(a_{1},a_{2},a_{3},a_{4}, a_{12}, d, \Delta)=(1, 0.5\mathrm{i}, 0.8\mathrm{i}, 0.7\mathrm{i}, 0.6, 0.3, 1)$, there exist two 3-bifurcation states denoted by $A$ and $B$ on the OBC spectrum. There exist two PBC spectra $E(|\beta|=\rho_{c})$ and $E(|\beta|=\rho^{-1}_{c})$ with $\rho_{c}=0.88$ passing through $A$. These two spectra exactly coincide with each other but are pointing in the opposite directions. What is particular in this model is that either of the spectra is composed of three closed loops, two of which are passing through $A$ and $B$ respectively, each three times, as can be seen from Fig.5(c). The circle $|\beta|=\rho_{c}$ ($|\beta|=\rho^{-1}_{c}$) corresponding to the PBC spectrum then intersects the GBZ at six points, which are the counterparts of the two 3-bifurcation states on the GBZ. In this model with $p=q=8$, for either of the two 3-bifurcation states, we have $|\beta_{6}|=|\beta_{7}|=|\beta_{8}|$ and $|\beta_{9}|=|\beta_{10}|=|\beta_{11}|$.
\nocite{*}
\bibliography{SMreference}